\documentclass[%
superscriptaddress,
twocolumn,
 amsmath,amssymb,
 aps,
floatfix,
]{revtex4-2}

\usepackage{graphicx}
\usepackage{dcolumn}
\usepackage{bm}
\usepackage{siunitx}
\usepackage{xcolor}

\begin{document}

\preprint{APS}

\title{Photoluminescence spectra of point defects in semiconductors: validation of first principles calculations
}

\author{Yu Jin}
\affiliation{Department of Chemistry, University of Chicago, Chicago, Illinois 60637, United States}
\author{Marco Govoni}
\email{mgovoni@anl.gov}
\affiliation{Pritzker School of Molecular Engineering, University of Chicago, Chicago, Illinois 60637, United States}
\affiliation{Materials Science Division and Center for Molecular Engineering, Argonne National Laboratory, Lemont, Illinois 60439, United States}
\author{Gary Wolfowicz}
\affiliation{Materials Science Division and Center for Molecular Engineering, Argonne National Laboratory, Lemont, Illinois 60439, United States}
\author{Sean E. Sullivan}
\affiliation{Materials Science Division and Center for Molecular Engineering, Argonne National Laboratory, Lemont, Illinois 60439, United States} 
\author{\\ F. Joseph Heremans}
\affiliation{Pritzker School of Molecular Engineering, University of Chicago, Chicago, Illinois 60637, United States}
\affiliation{Materials Science Division and Center for Molecular Engineering, Argonne National Laboratory, Lemont, Illinois 60439, United States}
\author{David D. Awschalom}
\affiliation{Pritzker School of Molecular Engineering, University of Chicago, Chicago, Illinois 60637, United States}
\affiliation{Materials Science Division and Center for Molecular Engineering, Argonne National Laboratory, Lemont, Illinois 60439, United States}
\affiliation{Department of Physics, University of Chicago, Chicago, Illinois 60637, United States}
\author{Giulia Galli}
\email{gagalli@uchicago.edu}
\affiliation{Department of Chemistry, University of Chicago, Chicago, Illinois 60637, United States}
\affiliation{Pritzker School of Molecular Engineering, University of Chicago, Chicago, Illinois 60637, United States}
\affiliation{Materials Science Division and Center for Molecular Engineering, Argonne National Laboratory, Lemont, Illinois 60439, United States}

\date{\today}

\begin{abstract}
Optically and magnetically active point defects in semiconductors are interesting platforms for the development of solid-state quantum technologies. Their optical properties are usually probed by measuring photoluminescence spectra, which provide information on excitation energies and on the interaction of electrons with lattice vibrations. We present a combined computational and experimental study of photoluminescence spectra of defects in diamond and SiC, aimed at assessing the validity of theoretical and numerical approximations used in first principles calculations, including the use of the Franck-Condon principle and the displaced harmonic oscillator approximation. We focus on prototypical examples of solid-state qubits, the divacancy centers in SiC and the nitrogen-vacancy in diamond, and we report computed photoluminescence spectra as a function of temperature that are in very good agreement with the measured ones. As expected we find that the use of hybrid functionals leads to more accurate results than semilocal functionals. Interestingly our calculations show that constrained density functional theory (CDFT) and time-dependent hybrid DFT perform equally well in describing the excited state potential energy surface of triplet states; our findings indicate that CDFT, a relatively cheap computational approach, is sufficiently accurate for the calculations of photoluminescence spectra of the defects studied here. Finally, we find that only by correcting for finite-size effects and extrapolating to the dilute limit, one can obtain a good  agreement between theory and experiment. Our results provide a detailed validation protocol of first principles calculations of photoluminescence spectra, necessary both for the interpretation of experiments and for robust predictions of the electronic properties of point defects in semiconductors.
\end{abstract}

\maketitle

\section{\label{sec:intro} Introduction}
The last two decades have witnessed the rapid development of quantum information technologies based on solid state platforms, in particular optically addressable spin-defects in semiconductors and insulators~\cite{heremans2016review, awschalom2018quantum, zhang2020material, wolfowicz2021quantumguideline}. The opto-electronic properties of point defects used to realize qubits are most often probed by measuring photoluminescence (PL) spectra, which yield information about excitation energies and the interaction of the excited electrons with lattice vibrations.

A PL experiment collects the photons emitted when an excited electron radiatively decays to the ground state (GS), and PL spectra of defects usually exhibit a narrow zero-phonon line (ZPL) and a broad phonon side band (PSB); the latter originates from decay processes that involve structural relaxation and thus the coupling of electrons and phonons. The strength of such coupling can be inferred from the Debye–Waller factor (DWF)~\cite{walker1979optical} that is proportional to the ratio between the emission intensity of the ZPL and that of the entire spectrum. Applications that require photon coherence or interference benefit from points defects whose PL signal exhibits a high DWF, indicating a weak coupling between phonons and electrons. The average number of phonons emitted during an electronic transition is instead quantified by the Huang-Rhys factor (HRF)~\cite{HR-theory,walker1979optical}, which can be estimated from measured spectra from the logarithm of the DWF.

As for many properties of condensed systems, Density Functional Theory (DFT) has turned out to be a valuable tool to compute PL spectra, which are used to interpret experiments as well as to provide predictions of the fingerprints of specific defects in materials~\cite{freysoldt2014RevModPhys,alkauskas2016tutorial,dreyer2018reviewarmr}. For example, first principles spectra based on DFT have been recently reported for nitrides, e.g., GaN~\cite{alkauskas2012plgan,reshchikov2021measurement}, AlN~\cite{aleksandrov2020luminescence}, and hexagonal born nitride (h-BN)~\cite{tawfik2017first, hamdi2020swhbn, grosso2020hbn, camphausen2020observation, jara2021carbonhbn, linderalv2021spehbn}, 
diamond~\cite{alkauskas2014luminescence,thiering2015sivhdiamond,norambuena2016siv,gali2017nvdjt, zhang2018nv0djt, gali2018groupiv, harris2020groupiii, karim2020nvpl, zemla2020sinvdiamond, razinkovas2020vibrational,karim2021bright}, silicon carbide (SiC)~\cite{udvarhelyi2020vsi, shang2020vsi, hassanzada20202dsic, arsalan2021vvpl}, and monolayers of transition metal dichalcogenides  (TMDC)~\cite{chakraborty2020dynamic}. These studies have been performed with several useful computational approaches; however, a systematic assessment of the theoretical and numerical approximations adopted in PL calculations has not yet been conducted.

In this work, we present a joint theoretical and experimental study of the PL spectra of prototypical spin-defects in diamond and SiC. We focus on the negatively charged nitrogen-vacancy (NV$^-$) in diamond~\cite{walker1979optical, kehayias2013nvexp, schirhagl2014nvsensor, doherty2013nitrogen, gali2019nvreview, su2019luminescence} and the neutral divacancy (V$_{\text{Si}}$V$_{\text{C}}^0$, abbreviated as VV$^0$) centers in hexagonal 4H-SiC which have been recently suggested as promising platforms for quantum sensors~\cite{wolfowicz2017optical, christle2017isolated, davidsson2018divacancy, son2020sic}. While the NV$^-$ center has just one possible geometrical configuration in the GS, the VV$^0$ centers may attain four different geometries due to the layered structure of 4H-SiC, giving rise to different PL signals. We discuss the comparison between theoretical and experimental PL spectra as a function of temperature, as well as HRFs and DWFs, and we present a detailed assessment of the theoretical and numerical approximations involved in first principles calculations. These approximations include the choice of the density functional, the method adopted to describe excited state (ES) potential energy surfaces, finite supercell size and approximations based on the Franck-Condon (FC) principle and the displaced harmonic oscillator (DHO) approximation.

We compare constrained DFT (CDFT)~\cite{gali2009nv} calculations of ES potential energy surfaces of triplet states with those carried out with time-dependent DFT (TDDFT), which enable the description of the ES as a linear combination of multiple Slater determinants. In our TDDFT calculations we use hybrid functionals with the fraction of exact exchange determined by the dielectric constant of the system; based on recent studies, these calculations are expected to yield results in good agreement with the solutions of the Bethe-Salpeter equation~\cite{sun2020BSETDDFT,sijia2021MLBSE}. So far, TDDFT has only been employed to describe spin defects with cluster models and atomic centered basis sets~\cite{gali2011tddft, petrone2016nvcluster, petrone2018sivtddft, karim2020nvpl, karim2021bright} and here we present a comparison between CDFT and TDDFT calculations carried out for the same supercell and using the same plane-wave basis sets and density functional. We also investigate finite size effects which affect defect-host interactions and present results for PL line shapes converged to supercells with more than 10,000 atoms, following the approach proposed by Alkauskas et al.~\cite{alkauskas2014luminescence, elisa2018siv, razinkovas2020vibrational}. Finally, we provide a qualitative assessment of the accuracy of the FC principle and the the displaced harmonic oscillator (DHO) approximation. By conducting fully converged hybrid functional calculations for the electronic properties, we obtain good agreement with measured spectra over a wide temperature range, with small differences between PL line shapes obtained with phonons computed with semilocal or hybrid functionals.

The rest of the paper is organized as follows. In Sec.~\ref{sec:theory} we discuss the methodology for computing PL spectra using the Huang-Rhys (HR) theory within the generating function formulation, highlighting all theoretical and numerical approximations. In Sec.~\ref{sec:tech-det}, we give the details of first-principles calculations and experiments. In Sec.~\ref{sec:results}, results on the chosen defects in SiC and diamond, including ZPLs, HRFs, DWFs and PL line shapes are discussed. Conclusions are given in Sec.~\ref{sec:conclusion}.

\section{\label{sec:theory}Theory and computational methodology}

\begin{figure}[t]
\includegraphics[width= 8.5 cm]{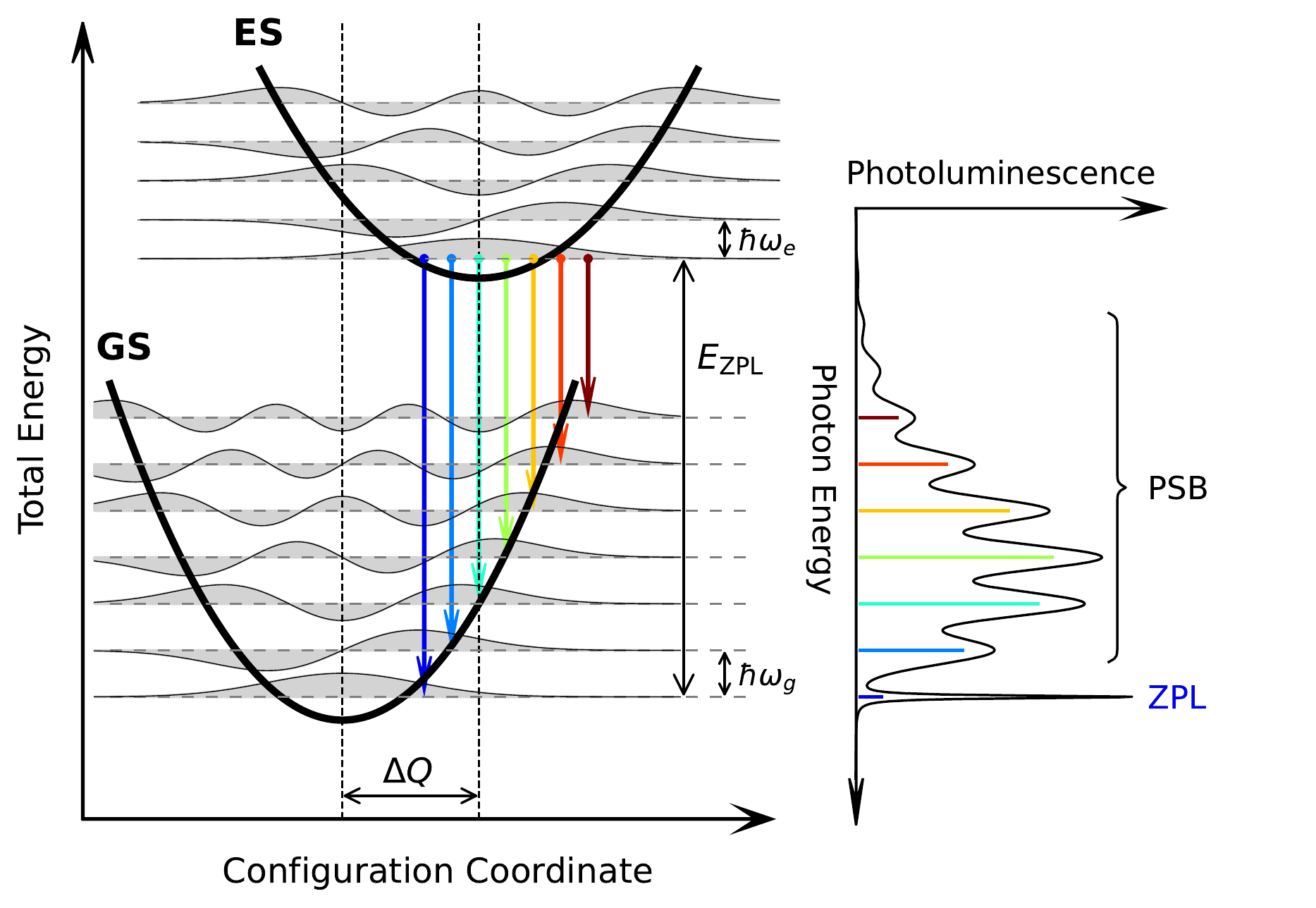}
\caption{\label{fig:schematic-pl} Schematic diagram illustrating optical processes leading to photoluminescence (PL) spectra. For simplicity only one phonon mode is depicted in the diagram. Ground state (GS) and excited state (ES) potential energy curves are approximated by harmonic functions with frequency $\omega_g$ and $\omega_e$, respectively. Vibrational energy levels and wavefunctions are shown as horizontal dashed lines and gray areas, respectively. $\Delta Q$ is the mass-weighted displacement between the local minimum of the GS and the ES energy curves. Colored arrows represent optical transitions at 0 K. The zero-phonon line (ZPL) originates from the transition between the $0$-th vibrational level of the ES to the $0$-th vibrational level of the GS. All other transitions contribute to the phonon sideband (PSB).}
\end{figure}


Based on Fermi's golden rule and the FC principle, the PL spectrum generated by the optical transitions from the ES to the GS potential energy surfaces, as a function of the photon energy $\hbar\omega$, is expressed as~\cite{alkauskas2014luminescence}:
\begin{equation}
\begin{aligned}
L(\hbar \omega, T)&=\frac{n\omega^{3}}{3 \pi \epsilon_{0} c^{3} \hbar}\left|\boldsymbol{\mu}_{eg} \right|^{2} \sum_{i} \sum_j P_{ej} (T) \left|\left\langle\Theta_{ej} \mid \Theta_{gi}\right\rangle\right|^{2}\\
&\times \delta\left(E_{\text{ZPL}}+E_{ej}-E_{g i} - \hbar \omega\right),
\end{aligned}
\label{eq:Ietog}
\end{equation}
where $\boldsymbol{\mu}_{eg}$ is the electronic transition dipole moment; $n$ is the refractive index of the material; $|\Theta^{gi}\rangle$  ($|\Theta^{ej}\rangle$) is the $i$-th ($j$-th) nuclear wavefunction of the system in the GS (ES) with vibrational energy $E_{gi}$ ($E_{ej}$); $E_{\text{ZPL}}$ is the energy of the ZPL (see Fig.~\ref{fig:schematic-pl}). The thermal distribution function of the vibrational energy in the ES is 
\begin{equation}
    P_{ej}(T) = \frac{e^{-\frac{E_{e j}}{k_BT}}}{\sum_j e^{-\frac{E_{e j}}{k_BT}}},
\end{equation}
where $k_B$ is the Boltzmann constant. For an ordered solid, under the harmonic approximation we express the nuclear wavefunctions as products of vibrational wavefunctions:
\begin{equation}
\label{eq:nuclear-wavefunctions}
    |\Theta_{gi}\rangle = \prod_{k} |\phi_{kn^{gi}_{k}}\rangle, \quad
    |\Theta_{ej}\rangle = \prod_{k} |\phi_{kn^{ej}_{k}}\rangle, 
\end{equation}
where $n^{gi}_{k}$ ($n^{ej}_{k}$) is the number of $k$-th phonons in the $i$-th ($j$-th) vibrational state of the GS (ES), and $|\phi_{kn}\rangle$ is the $n$-th excited state of the $k$-th phonon mode. The vibrational energies in the GS and ES are: 
\begin{equation}
    \label{eq:egi-eej}
    E_{g i} = \sum_{k} n^{gi}_{k}\hbar\omega^{g}_{k},
     \quad
    E_{e j} = \sum_{k} n^{ej}_{k}\hbar\omega^{e}_{k},
\end{equation}
where $\omega^g_k$ ($\omega^e_k$) is the frequency of $k$-th phonon in the GS (ES). Note that by definition Eq.~\ref{eq:egi-eej} does not include the zero point energy which is included in the term $E_{\text{ZPL}}$.

A commonly used approximation is the so called displaced harmonic oscillator (DHO) approximation, (or HR theory \cite{HR-theory}), where Eq.~\ref{eq:Ietog} is simplified by assuming that the potential energy surface of the ES and the GS are identical except for a rigid displacement due to the difference in their equilibrium structures, i.e., $\omega_k^g = \omega_k^e$ (the superscript $g$ and $e$ will hence be omitted), and:
\begin{equation}
\begin{aligned}
    &\left|\left\langle\Theta_{ej} \mid \Theta_{gi}\right\rangle\right|^{2} \\
    &\quad=
    \prod_k  e^{-S_k} (S_k)^{n_k^{gi}-n_k^{ej}} \left(\frac{n_k^{ej} !}{n_k^{gi} !}\right)\left[L_{n_k^{ej}}^{n_k^{gi}-n_k^{ej}}(S_k)\right]^{2},
    \label{eq:DHOoverlap}
\end{aligned}
\end{equation}
where $S_k$ is the partial HRF and accounts for the average number of $k$-th phonons that participate in the transition. $L_m^{n-m}$ are the associated Laguerre polynomials~\cite{laguerre-poly}. For the calculations of PL line shapes, only the phonons of the GS are computed and used within the DHO approximation. At zero temperature, Eq.~\ref{eq:Ietog} may be further approximated by setting to zero the vibrational energy in the ES (i.e., we only consider $j=0$, and $n^{e0}_k=0$), namely: 
\begin{equation}
\begin{aligned}
    &L(\hbar \omega, T=0) \\
    &\quad\propto \omega^3 \sum_{i} 
    \left[\prod_k\frac{e^{-S_k}}{n^{gi}_k!}(S_k)^{n^{gi}_k}\right]
    \delta\left(E_{Z P L}-E_{gi} - \hbar \omega\right).
    \label{eq:LDHOzeroT}
\end{aligned}
\end{equation}

To avoid the evaluation of the overlap integrals entering Eq.~\ref{eq:DHOoverlap} and the sum over all vibrational states of the GS and the ES, we adopt the generating function approach~\cite{lax1952generatingfxn, kubo1955generatingfxn} to compute PL spectra. In the DHO approximation Eq.~\ref{eq:Ietog} can be obtained as the Fourier transform of the generating function $G(t,T)$:
\begin{equation}
    L(\hbar\omega, T) \propto \omega^3 \int_{-\infty}^{\infty} \mathrm{d}t G(t,T) e^{i\omega t - \frac{\lambda}{\hbar} |t| -i\frac{E_{\text{ZPL}}}{\hbar }t  },
    \label{eq:Lgenfun}
\end{equation}
where
\begin{equation}
\begin{aligned}
    G(t,T) = \exp \bigg\{&- \sum_k S_{k} \left[ \left( 1 - e^{i\omega_kt} \right)\right. \\
    &\left.+ \bar{n}_k(T) \left( 2 - e^{i\omega_kt} - e^{-i\omega_kt} \right) \right]\bigg\},
    \label{eq:genfunct}
\end{aligned}
\end{equation}
and $\lambda$ accounts for the broadening of the line shape. $\bar{n}_k (T)$ is the average occupation number of the $k$-th phonon mode:
\begin{equation}
    \bar{n}_k (T) = \dfrac{1}{e^{\frac{\hbar\omega_k}{k_BT}} - 1}.
\end{equation}
In practice, one may write Eq.~\ref{eq:genfunct} as the following alternative expression: 
\begin{equation}
    G(t,T) = e^{S(t) - S(0) + C(t,T) + C(-t,T) - 2C(0,T)},
\end{equation}
where $S(t)=\sum_k S_k e^{i\omega_kt}$ and $C(t,T)= \sum_k \bar{n}_k (T) S_k e^{i\omega_kt}$ are evaluated as the Fourier transforms of the spectral densities:
\begin{equation}
\label{eqn:e-ph-spectral}
    S(\hbar\omega) = \sum_k S_k \delta\left( \hbar\omega - \hbar\omega_k \right),
\end{equation}
\begin{equation}
\label{eqn:temp-spectral}
    C(\hbar\omega, T) = \sum_k \bar{n}_k (T) S_k \delta (\hbar\omega - \hbar\omega_k).
\end{equation}
In Eqs.~\ref{eqn:e-ph-spectral}-\ref{eqn:temp-spectral}, the $\delta$-functions are replaced by Gaussian functions with $\omega$-dependent broadening to account for a continuum of vibrational modes participating in the optical transition (see Sec.~\ref{subsec:hrf}).

At zero temperature, we have
$\bar{n}_k (T=0) \approx 0$ and $C(\hbar\omega,T=0) \approx 0$, and Eq.~\ref{eq:Lgenfun} is equivalent to Eq.~\ref{eq:LDHOzeroT} with the $\delta$-function replaced by a Lorentzian function with a broadening $\lambda$. In order to evaluate Eq.~\ref{eq:Lgenfun}, the partial HRF $S_k$, $E_{\text{ZPL}}$, and phonon frequencies $\omega_k$ are required as input. We compute $S_k$ as:
\begin{equation}
    \label{eq:Sk}
    S_k = \dfrac{\omega_k \Delta Q_k^2}{2\hbar},
\end{equation}
where $\Delta Q_k$ is the mass-weighted displacement along the $k$-th mode, evaluated as:
\begin{equation}
    \label{eqn:r-dq}
    \Delta Q_{k}=\sum_{\alpha=1}^N\sum_{i=x,y,z} \sqrt{M_{\alpha}} \Delta \mathbf{R}_{\alpha i} \boldsymbol{e}_{k, \alpha i}.
\end{equation}
In Eq.~\ref{eqn:r-dq}, $\mathbf{e}_{k,\alpha i}$ is the eigenvector of the $k$-th phonon mode on the $\alpha$-th atom in the $i$-th direction; $M_{\alpha}$ is the mass of the $\alpha$-th atom, and $\Delta \mathbf{R}_{\alpha i} = (\mathbf{R}_{\alpha i})_e - (\mathbf{R}_{\alpha i})_g $ is the displacement between the ES and the GS equilibrium atomic structures in the $i$-th direction. Within the harmonic approximation, $\Delta Q_k$ may be equivalently computed as \cite{alkauskas2014luminescence}:
\begin{equation}
    \label{eqn:f-dq}
    \Delta Q_{k}=\frac{1}{\omega_{k}^{2}} \sum_{\alpha=1}^N \sum_{i=x,y,z} \frac{\mathbf{F}_{\alpha i}}{\sqrt{M_{\alpha}}} \boldsymbol{e}_{k, \alpha i}.
\end{equation}
Here $\mathbf{F}_{\alpha i}$ is the GS force on the $\alpha$-th atom in the $i$-th direction evaluated at the ES equilibrium structure.

In this work, we simulate PL spectra at finite temperature using Eq.~\ref{eq:Lgenfun}, with parameters computed from first principles. In particular, we use DFT to obtain the GS equilibrium atomic structure, $\mathbf{R}_g$, the GS forces, $\mathbf{F}$, the phonon modes, $\mathbf{e}$, and the phonon frequencies $\omega$; the ES equilibrium atomic structure, $\mathbf{R}_e$, is obtained with CDFT and, in some cases validated by carrying out TDDFT calculations. The resulting PL line shapes are then compared with measured PL spectra at finite temperature. Below, we report our results and we systematically investigate the validity of the chosen theoretical and numerical approximations, including the use of the FC and the DHO approximations, which are at the core of the HR theory.

\section{\label{sec:tech-det} Technical details}
\subsection{\label{sub-sec:fp-calculations}First principles calculations}
The electronic structures of the defects in diamond and 4H-SiC are obtained using DFT and the planewave pseudopotential method, as implemented in the Quantum Espresso package~\cite{QE-2009, QE-2017, QE-exascale}. The planewave energy cutoff was set to 80 Ry. We used SG15 ONCV norm-conserving pseudopotentials~\cite{ONCV_1, SCHLIPF201536_ONCV_2} and the semi-local functional by Perdew, Burke, and Ernzerhof (PBE)~\cite{PBE}, the dielectric dependent hybrid (DDH) functional~\cite{skone2014ddh} and the screened hybrid functional by Heyd, Scuseria, and Ernzerhof (HSE)~\cite{HSE03,HSE06}. The fraction of exact exchange used in the DDH functional was determined by the inverse of the macroscopic dielectric constant of the system, resulting in 18\% and 15\% of exact exchange for diamond and 4H-SiC~\cite{skone2014ddh,hosung2017qubit}, respectively. The macroscopic dielectric constants were computed by including the full response of the electronic density to the perturbing external electric field at the level of hybrid functional, and the fraction of exact exchange was self-consistently determined from the dielectric constant~\cite{skone2014ddh}.

We used a $(4\times 4\times 4)$ supercell with 512 atomic sites and a $(5\times 5\times 2)$ supercell with 400 atomic sites for the NV$^-$ center in diamond and VV$^0$ centers in 4H-SiC, respectively. In the cases of VV$^0$ centers in 4H-SiC, convergence tests were carried out with large supercells (up to $(8\times 8\times 2)$). We used the lattice constant optimized with each specific functional (see Tab.~S1 of the Supplementary Information (SI)~\cite{SI}). The Brillouin zone was sampled with the $\Gamma$ point.

The paramagnetic ESs were computed using the CDFT (also called $\Delta$SCF) method, where one electron is promoted from the highest occupied to the lowest unoccupied state in the same spin channel (see Sec.~II of the SI for details of CDFT calculations). The energy $E_{\text{ZPL}}$ was computed as the difference of the total energy of the relaxed ES (with CDFT) and that of the GS. The CDFT method has been shown to yield reliable results for systems with localized electronic states, e.g., the NV$^-$ center in diamond~\cite{gali2009nv} and VV$^0$ centers in 4H-SiC~\cite{gordon2015vvsic, davidsson2018divacancy, arsalan2021vvpl}. We also performed TDDFT calculations within the Tamm-Dancoff approximation to assess the accuracy of our CDFT results. We obtained the ES low-lying energies and eigenvectors by iteratively diagonalizing the linearized Liouville operator, as implemented in the WEST code~\cite{govoni2015west, nguyen2019ffbse}. Due to the higher computational cost of TDDFT calculations, we used a $(3\times 3\times 3)$ supercell and a $(5\times 3 \sqrt{3} \times 1)$ supercell for the NV$^-$ center in diamond and VV$^0$ centers in 4H-SiC, respectively.

Phonon modes of bulk and defective solids were computed using the frozen phonon approach, with configurations generated with the PHONOPY package~\cite{phonopy} and a displacement of 0.01 \AA{} from equilibrium structures. Phonon calculations for pristine bulk systems were carried out with the PBE, DDH and HSE functionals, but those for defective solids were performed only with PBE due to the high computational cost of hybrid functionals. We then approximated the values of hybrid-DFT phonons by using a scaling factor equal to the ratio of hybrid-DFT and PBE phonon results in the pristine bulk systems. We verified that this approximation yields accurate phonon frequencies for bulk systems (see Sec.~IV of the SI for details). We evaluated finite size effects on computed PL line shapes, HRFs and spectral densities following the force constant matrix embedding approach proposed by Alkauskas et al.~\cite{alkauskas2014luminescence, elisa2018siv, razinkovas2020vibrational} (see Sec.~V of the SI).

\subsection{\label{sub-sec:expt.}Experiments}
The SiC experiments were realized in a confocal microscopy setup (0.67 NA objective) with the sample in a closed-cycle cryostat at 10 K, unless mentioned otherwise. The sample was diced from a commercial high-purity semi-insulating 4H-SiC wafer (Cree) containing intrinsic concentrations of VV$^0$ (10$^{15}$-$10^{16}$ cm$^{-3}$). The sample was excited within the VV$^0$ absorption sideband with a 908 nm laser (QPhotonics, $\sim$100 mW), and the resulting PL was filtered using several 1000 nm long-pass filters. The PL was then measured using a spectrometer with a 300 g/mm grating blazed for \SI{1.2}{\micro\meter} and an InGaAs camera (Teledyne Princeton Instruments) with a spectral resolution of $\sim$0.3 nm. Careful calibration was performed to correct for the entire setup transmission and the camera response using a NIST calibrated tungsten-halogen white light source (StellarNet). Optically-detected magnetic resonance at a weak static magnetic field ($<$20 G) with microwave excitation delivered using a printed circuit board was used to obtain independent contrast (i.e., the difference in PL with and without microwave excitation) for the various VV$^0$ sites. The DWF is calculated as the ratio of the integrated intensity of the ZPL to the total integrated intensity. Though most of the PL is within our detection bandwidth (900-1600 nm), the sideband may be very slightly underestimated due to weak emission extending beyond this range.

Similarly, for the NV$^-$ center in diamond, the PL spectra were taken on an ensemble of NV$^-$ centers using a home-built confocal microscope with a flow cryostat (Janis - LakeShore Cryotronics) for temperature studies.  The sample was a IB diamond (Sumitomo) with a high nitrogen concentration synthesized via high-pressure / high-temperature growth. The sample was electron irradiated (2 MeV, 10$^{17}$ cm$^{-2}$) and annealed (850 \textdegree C, 2 hr) resulting in a high density of NV$^-$ centers. The NV$^-$ center ensemble was photo-excited within the absorption sideband using 532 nm light and the PL measurements were collected using a high resolution spectrometer with a visible camera (Teledyne Princeton Instruments) with $\sim$0.1 nm spectral resolution. The spectrum intensity was also calibrated using a tungsten halogen light source (Ocean Optics) to correct for transmission losses in the experimental set-up.

\section{\label{sec:results}Results and Discussion}

\subsection{\label{subsec:zpl}Zero-phonon lines}

\begin{figure}[b]
\includegraphics[width= 8.5 cm]{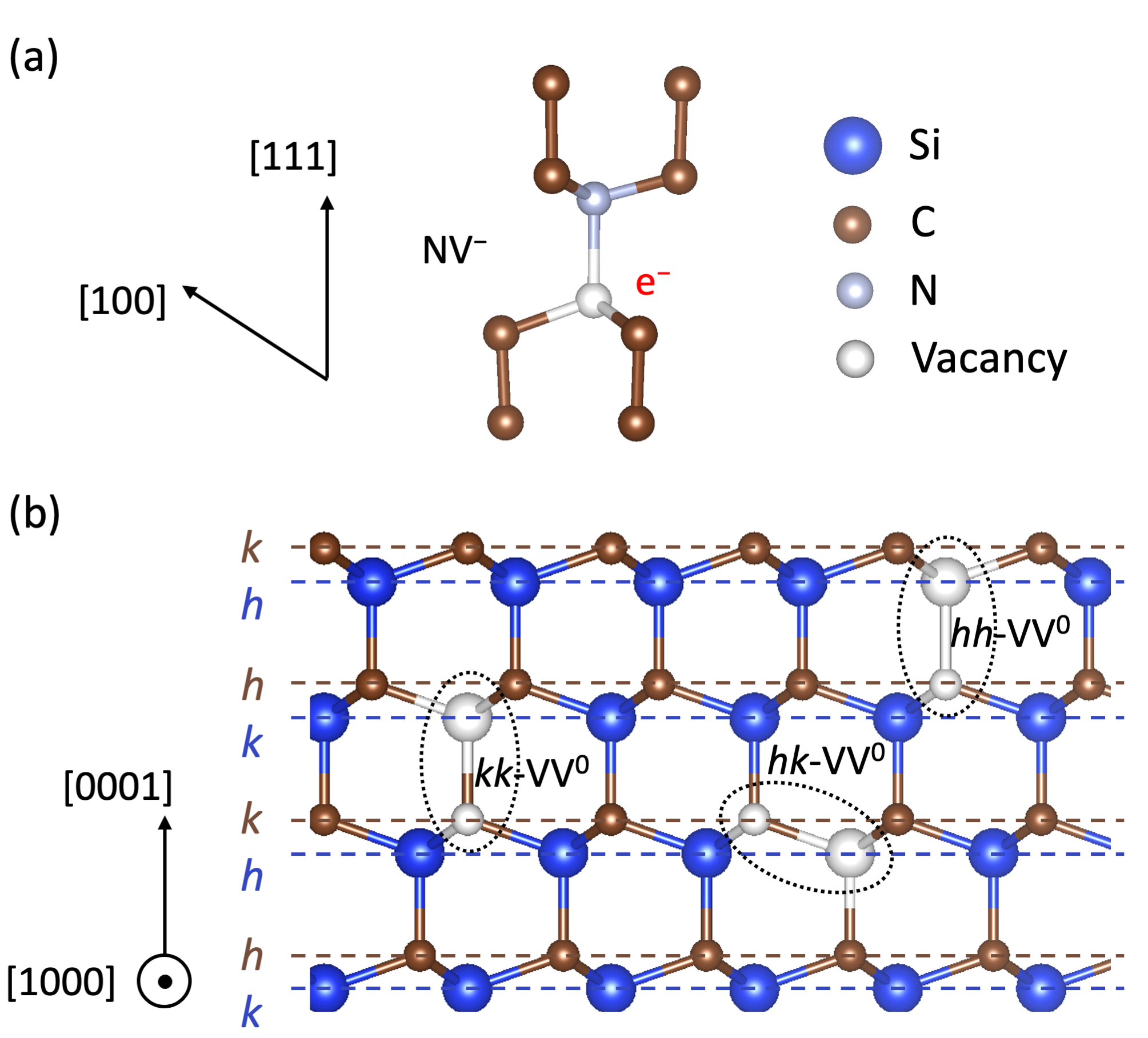}
\caption{\label{fig:defect-geo} Ball and stick representation of the defect centers studied in this work. (a) NV$^-$ center in diamond. (b) Divacancy ($\text{V}_{\text{Si}}\text{V}_{\text{C}}^0$) centers in 4H-SiC. The planes are labelled with $h$ and $k$ according to the symmetry of lattice sites. Three non-equivalent configurations ($hh$, $kk$ and $hk$) of the  $\text{V}_{\text{Si}}\text{V}_{\text{C}}^0$ centers are shown. The $kh$ configuration was not investigated computationally, due to the lower quality of the experimental spectrum for this configuration.}
\end{figure}

\begin{figure*}
\includegraphics[width=16 cm]{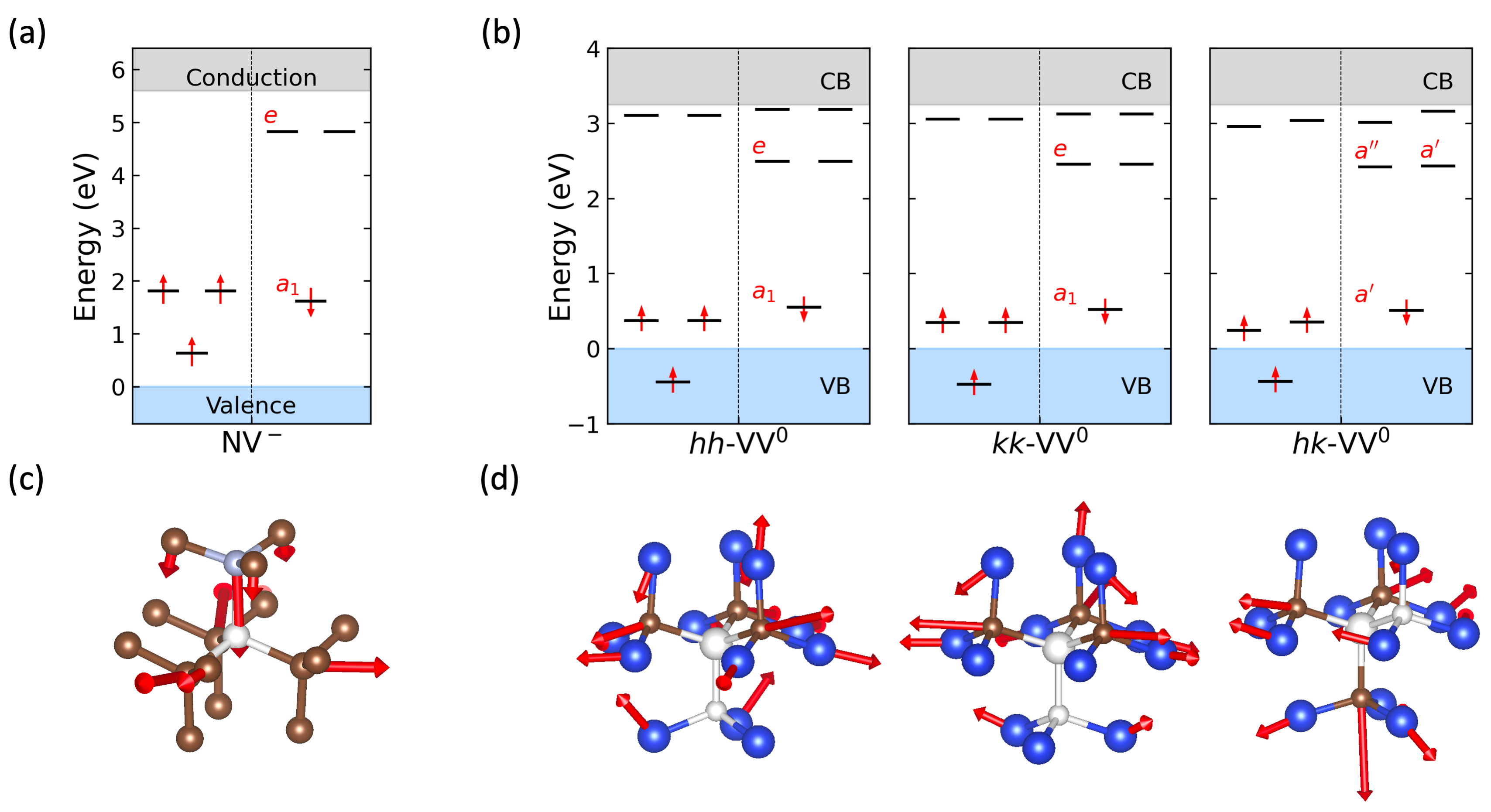}
\caption{\label{fig:defect-levels}Ground state (GS) electronic structure of (a) NV$^-$ center in diamond and (b) VV$^0$ centers in 4H-SiC at the DDH level of theory. Displacements between the excited state (ES) and the GS equilibrium structures of the NV$^-$ center in diamond and the VV$^0$ centers in 4H-SiC are shown in (c) and (d), respectively. The red arrows represent mass-weighted displacements ($\Delta Q$) of each atom with the magnitude being amplified by a factor of 10. The $\Delta Q$ of the NV$^-$ center is mainly localized on the nitrogen atom and three nearest neighbor carbon atoms around the carbon vacancy. The $\Delta Q$ of the VV$^0$ centers is mainly localized on three nearest neighbor carbon atoms and nine next nearest neighbor silicon atoms around the silicon vacancy, as well as three nearest neighbor silicon atoms around the carbon vacancy.}
\end{figure*}

The NV$^-$ center in diamond is composed of a nitrogen impurity and an adjacent carbon vacancy (V$_{\mathrm{C}}$) with an additional electron, as shown in Fig.~\ref{fig:defect-geo}(a)~\cite{doherty2013nitrogen,gali2019nvreview}. It has $C_{3v}$ symmetry, and three defect orbitals are present within the band gap of diamond: the $a_1$ orbital and the two-fold degenerate $e$ orbitals, as shown in Fig.~\ref{fig:defect-levels}(a). Defect orbitals are mainly localized on three carbon atoms around the $\text{V}_{\mathrm{C}}$ (see Fig.~S1 of the SI).

The neutral divacancy center in 4H-SiC is composed of a silicon vacancy (V$_{\mathrm{Si}}$) and an adjacent carbon vacancy (V$_{\mathrm{C}}$) and is denoted as V$_{\mathrm{Si}}$V$_{\mathrm{C}}^0$~\cite{christle2017isolated,wolfowicz2017optical}. We consider the 4H polytype of SiC, 4H-SiC, with ABCB stacking along the c-axis. 4H-SiC contains two nonequivalent hexagonal ($h$) and quasi-cubic ($k$) sites for each type of atom, as shown in Fig.~\ref{fig:defect-geo}(b). Therefore, the VV$^0$ center can occur in four distinct configurations ($hh$, $kk$, $hk$ and $kh$), and the experimentally measured PL ensemble is a mixture of contributions from all configurations. Experimentally, we used microwave-assisted spectroscopy to separate the PL of different configurations. In our computational study, we considered the $hh$, $kk$, and $hk$ configurations denoted as $hh$-VV$^0$, $kk$-VV$^0$ and $hk$-VV$^0$.
The first two $c$-axis orientated defects ($hh$-VV$^0$ and $kk$-VV$^0$) have $C_{3v}$ symmetry, with an $a_1$ state and two sets of degenerate $e$ states within the band gap of 4H-SiC (see Fig.~\ref{fig:defect-levels}(b)). The $a_1$ orbital and lower $e$ orbitals are mainly localized on three carbon atoms around the $\text{V}_{\mathrm{Si}}$ (see Fig.~S1 of the SI). For NV$^-$, $hh$-VV$^0$ and $kk$-VV$^0$, we studied the optical transition between the $a_1$ orbital and the (lower) $e$ orbitals, which corresponds to the transition between the $^3A_2$ and the $^3E$ many-body states. For the $hk$ configuration, which  has $C_{1h}$ symmetry, we studied the $a^\prime$ and the $a^{\prime\prime}$ transition accounting for the transition between the $^3A^{\prime\prime}$ and the $^3A^{\prime}$ state.

We computed $E_{\text{ZPL}}$ using the PBE, DDH, and HSE functionals in the DHO approximation (note that the zero-point energy contributions of the GS and ES phonons cancel out within the DHO approximation). Triplet ESs were computed using CDFT with electronic configuration $a_1^1e_x^2e_y^1$ (see Sec.~II of the SI for details of CDFT calculations). Results for all defect systems at different levels of theory are summarized in Tab.~\ref{tab:zpl}. The PBE functional underestimates the measured $E_{\text{ZPL}}$ of the NV$^-$ by 0.24 eV, while the DDH and HSE functionals overestimate it by 0.26 eV. In the case of VV$^0$ centers, particular attention must be exercised to account for finite size effects. For small cells (e.g. $(5\times5\times2)$ supercell), the experimental order of the ZPL of the various defect configurations is not reproduced. Hence we computed $E_\text{ZPL}$ at the PBE level of theory for $(5\times5\times2)$ and $(8\times8\times2)$ supercells; we then used the difference between these two values to estimate the converged hybrid-DFT $E_\text{ZPL}$ starting from our results obtained with $(5\times5\times2)$ supercells. The converged results reproduce the experimental trend. Similar to our results for diamond, we find that the PBE functional underestimates the measured $E_{\text{ZPL}}$ of the VV$^0$ centers by $\sim$0.16 eV,  while the DDH (HSE) yields an overestimate of  $\sim$0.11 eV ($\sim$0.12 eV). Tab.~\ref{tab:zpl} summarizes our results and previously reported theoretical predictions of $E_{\text{ZPL}}$. We note that although the $E_{\text{ZPL}}$ obtained with the HSE functional is generally in good agreement with experiments, theoretical results exhibit a variance up to 0.3 eV due to different choices of pseudopotentials, supercell sizes and sampling of the reciprocal space.

\begin{table*}
\caption{\label{tab:zpl}Energy of the zero-phonon line ($E_{\text{ZPL}}$, eV) for spin-conserving transitions computed using different levels of theory and a $(4\times4\times4)$ supercell for the NV$^-$ center in diamond, and a $(5\times 5 \times 2)$ supercell for the VV$^0$ centers in 4H-SiC. The extrapolation value of the $E_{\text{ZPL}}$ with the finite size corrections is reported in parentheses. The finite size corrections are calculated as the difference of $E_{\text{ZPL}}$ values between the $(8\times 8 \times 2)$ and the $(5\times 5 \times 2)$ supercell at the PBE level of theory. Previous theoretical predictions on $E_{\text{ZPL}}$ are also shown.}
\begin{ruledtabular}
\begin{tabular}{c|c|c|c|c|c|c|c|c}
    Hosts & Defects &\multicolumn{3}{c|}{This work}& \multicolumn{3}{c|}{Previous Theoretical Work}& Expt.\\\cline{3-8}
    && PBE & DDH & HSE &PBE & DDH & HSE&   \\
    \hline
    Diamond & NV$^{-}$ &1.706 &2.205 &2.205 &1.72\footnote{Ref.~\cite{hosung2017qubit}: Calculations were carried out using the Quantum Espresso package with ONCV pseudopotentials (PPs). The planewave energy cutoff was set to 75 Ry. The Brillouin zone was sampled with the $\Gamma$ point. A $(4\times4\times4)$ ($(5\times3\sqrt{3}\times1)$) supercell was used for NV$^-$ ($hh$-VV$^0$).}, 1.706\footnote{Ref.~\cite{gali2009nv}: Calculations were carried out using the VASP code with PAW PPs. The planewave energy cutoff was set to 420 eV. The Brillouin zone was sampled with the $\Gamma$ point. A $(4\times4\times4)$ supercell was used for NV$^-$.} &2.22$^{\text{a}}$ &2.23$^{\text{a}}$, 1.955$^{\text{b}}$&1.945~\cite{doherty2013nitrogen}\\
    \hline
    4H-SiC & $hh$-VV$^0$ &1.086 (0.937) &1.346 (1.196) &1.371 (1.221) &1.03$^{\text{a}}$, 0.92\footnote{Ref.~\cite{davidsson2018divacancy}: Calculations were carried out using the VASP code with PAW PPs. The Brillouin zone was sampled with the $\Gamma$ point. A $(10\times10\times3)$ ($(8\times8\times3)$) supercell was used for PBE (HSE) calculations. HSE $E_{\text{ZPL}}$ was computed with PBE structure.} &1.30$^{\text{a}}$ &1.33$^{\text{a}}$, 1.056$^{\text{c}}$, 1.13\footnote{Ref.~\cite{gordon2015vvsic}: Calculations were carried out using the VASP code with PAW PPs. The planewave energy cutoff was set to 400 eV. The Brillouin zone was sampled using a $2\times2\times2$ k-point mesh. A $(4\times3\times1)$ supercell was used for VV$^0$.}, 1.14\footnote{Ref.~\cite{arsalan2021vvpl}: Calculations were carried out using the VASP code with PAW PPs. The planewave energy cutoff was set to 400 eV. The Brillouin zone was sampled with the $\Gamma$ point. A $(5\times5\times2)$ supercell was used for VV$^0$. HSE $E_{\text{ZPL}}$ was computed with PBEsol structure.}&1.095 \\
    & $kk$-VV$^0$ &1.105 (0.951) &1.355 (1.201) &1.372 (1.218) &0.94$^{\text{c}}$& &1.044$^{\text{c}}$, 1.14$^{\text{d}}$&1.096 \\
    & $hk$-VV$^0$ &1.075 (0.979) &1.355 (1.259) &1.365 (1.269) &0.97$^{\text{c}}$ & &1.103$^{\text{c}}$, 1.21$^{\text{d}}$&1.149 \\
\end{tabular}
\end{ruledtabular}
\end{table*}

\begin{figure}[b]
\includegraphics[width= 7 cm]{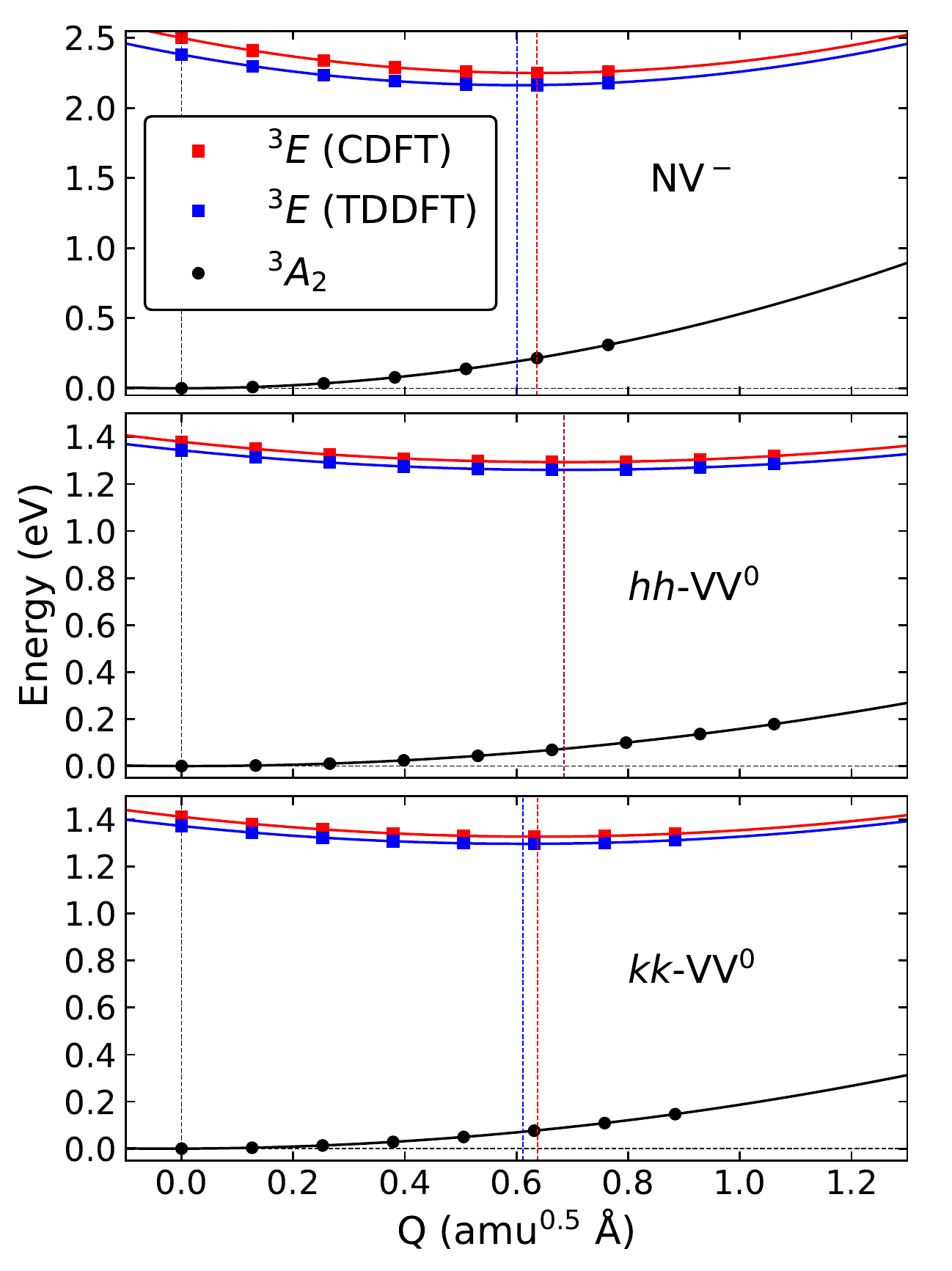}
\caption{\label{fig:cc-diagram} Configuration coordinate diagrams describing the total energies of the $^3A_2$ ground state (GS) and the $^3E$ excited state (ES) along the relaxation path resulting from CDFT with electronic configuration $a_1^1e_x^{1.5}e_y^{1.5}$ for the NV$^-$ center in diamond and the $hh$-VV$^0$ and the $kk$-VV$^0$ centers in 4H-SiC. Calculations are performed at the DDH level of theory. Dashed vertical lines denote the locations of the local minimum by fitting the energy curves with quadratic functions. The energy of the effective phonon obtained from the fitting process is 66.4 meV for the $^3A_2$ state and 72.0 (71.1) meV for the $^3E$ state, when computed with CDFT (TDDFT) for NV$^-$. The energy of the effective phonon is 36.8 meV for the $^3A_2$ state and 39.3 (38.6) meV for the $^3E$ state, when computed with CDFT (TDDFT) for $hh$-VV$^0$. The energy of the effective phonon is 38.7 meV for the $^3A_2$ state and 41.7 (41.2) meV for the $^3E$ state, when computed with CDFT (TDDFT) for $kk$-VV$^0$.}
\end{figure}

To estimate the accuracy of CDFT for the ES potential energy surface, we compared results obtained with CDFT and TDDFT for NV$^-$, $hh$-VV$^0$, and $kk$-VV$^0$ at the DDH level of theory. TDDFT enables the description of the ES as a linear combination of multiple Slater determinants. In our TDDFT calculations we use hybrid functionals with the fraction of exact exchange determined by the dielectric constant of the system; based on recent studies, these calculations are expected to yield results in good agreement with the solutions of the Bethe Salpeter equation~\cite{sun2020BSETDDFT,sijia2021MLBSE}. Previous TDDFT studies of these defect systems used cluster models and atomic-centered basis sets~\cite{gali2011tddft, petrone2016nvcluster, karim2020nvpl}. Here we consistently compared TDDFT and CDFT calculations performed with the same basis set and pseudopotential, in the same cell and with the same functional. After obtaining the equilibrium structure of the ES using the CDFT approach, we selected several configurations along the linear path connecting equilibrium atomic structures of the GS and the ES, and carried out single-point DFT, CDFT, and TDDFT calculations, from which configuration coordinate diagrams were obtained, as shown in Fig.~\ref{fig:cc-diagram}. The comparison between TDDFT and CDFT results was carried out using the configuration $a_1^1e_x^{1.5}e_y^{1.5}$ in CDFT because we could not converge the conﬁguration $a_1^1e_x^2e_y^1$ when using the DDH functional. The difference between CDFT calculations at the PBE level using the $a_1^1e_x^2e_y^1$ and $a_1^1e_x^{1.5}e_y^{1.5}$ electronic configurations is $\sim$0.04 eV; assuming an energy difference of the same order of magnitude at the DDH level, we deemed the CDFT/TDDFT comparison with the $a_1^1e_x^{1.5}e_y^{1.5}$ configuration to be a meaningful one. We found that the TDDFT energies of the $^3E$ states are 0.1 (0.04) eV smaller than CDFT energies for NV$^-$ (VV$^0$); the minimum of the TDDFT curve is close to that of the CDFT curve, with a difference smaller than 6\% for all three systems. An analysis of our TDDFT results shows that the spin-conserving transition between the $a_1$ and the $e$ orbitals contributes by more than 95\% to the whole transition to the $^3E$ state, indicating that the $^3E$ state can be well-described by a single Slater determinant. This finding is in good agreement with that of a previous TDDFT study using a cluster model, atom-centered basis sets and the PBE0 functional~\cite{gali2011tddft}. We note that the quality of the agreement between TDDFT and CDFT depends on the functional and the system. For example, we found a difference of $\sim$0.2 eV in the excitation energy and a difference greater than 12\% in the configuration coordinate of the minimum between TDDFT and CDFT results when using the PBE functional (see Fig.~S3 of the SI). In addition, in the case of defect systems with singlet GS and ES, we expect non-negligible differences between TDDFT and CDFT results, since the accurate description of a singlet ES requires a linear combination of at least two Slater determinants~\cite{MS2019carbondimerhbn, hamdi2020swhbn}. In that case the use of quantum embedding theories (QDET), should be preferable to describe strongly correlated states~\cite{he2020npj,he2020pccp,he2021jctc}.

To understand the effect of lattice vibrations on optical transitions, we analyzed in detail the mass-weighted displacement $\Delta Q$ computed at different levels of theory (see Sec.~III of the SI). The magnitude of the displacement follows the relation: HSE $>$ DDH $>$ PBE and the difference of the results obtained with different functionals can be up to 10\%. The same trend can also be found for energies of bulk phonons, 
and can be understood by noting that the bonds in the diamond and SiC crystals turn out to be stiffer with HSE than with DDH, which are in turn stiffer than with PBE, as reflected in the difference of predicted lattice constants (see Tab.~S1 of the SI). $\Delta Q$ is localized on the neighboring atoms of the defect center, as shown in Fig.~\ref{fig:defect-levels}(c) and (d), consistent with the localization of defect orbitals involved in the optical transition. We also found that the symmetry of the ES is reduced from $C_{3v}$ to $C_{1h}$ for NV$^-$, $hh$-VV$^0$ and $kk$-VV$^0$, and hence in principle both $a_1$ type and $e$ type phonons could participate in the optical process. Although the average symmetry of the ES structure turns out to be $C_{3v}$ due to the dynamic Jahn-Teller effect~\cite{abtew2011nvdjt, gali2017nvdjt}, here we used the ES equilibrium structure with $C_{1h}$ symmetry to include the coupling with $e$ type phonons. Hence the ES configuration used is $a_1^1e_x^2e_y^1$.

\subsection{\label{subsec:hrf}Huang-Rhys factor and spectral density of the electron-phonon coupling}

Tab.~\ref{tab:hrf} summarizes HRFs computed using Eq.~\ref{eq:Sk} at different levels of theory and includes combinations of mass-weighted displacements, $\Delta Q_k$, computed with DDH or HSE and phonons computed with PBE. We used a scaling factor  (see Sec.~IV of the SI) to approximate DDH (HSE) phonon frequencies using PBE results; the factor was evaluated from the ratio of frequencies of bulk systems optical phonons computed at different levels of theory, an approximation that introduces a root-mean-square error (RMSE) of only 0.4 meV for bulk 4H-SiC.

\begin{table*}
  \caption{Huang-Rhys factors (HRFs) for spin-conserving transitions computed using Eq.~\ref{eq:Sk} with different levels of theory. PBE$-\Delta Q$ denotes that $\Delta Q$ used in Eq.~\ref{eq:Sk} is computed using Eq.~\ref{eqn:f-dq} with forces and phonons computed at the PBE level. PBE$-ph$ denotes that phonons used in Eq.~\ref{eq:Sk} are computed at the PBE level. Similar notations are used for DDH and HSE. Only one digit was kept for HRFs computed with the largest supercells considering the uncertainty introduced in the extrapolation procedure (see Sec.~V of the SI). Experimental HRFs for VV$^0$ centers are estimated as the negative logarithm of the Debye-Waller factor (DWF).}
  \label{tab:hrf}
  \begin{ruledtabular}
  \begin{tabular}{c|c|c|c|cc|cc|c}
    Hosts&Defects&Cell Size&\multicolumn{1}{c}{PBE$-\Delta Q$}&\multicolumn{2}{c}{DDH$-\Delta Q$}&\multicolumn{2}{c}{HSE$-\Delta Q$}&Expt.\\
    &&&PBE$-ph$&PBE$-ph$&DDH$-ph$&PBE$-ph$&HSE$-ph$& \\
    \hline
    Diamond& NV$^{-}$ &$\left(4 \times 4 \times 4\right)$ &2.94 &3.20&3.32 &3.46 &3.64 &3.49 \cite{kehayias2013nvexp} \\
    &&$\left(12 \times 12 \times 12\right)$ &3.0 &3.2 &3.3 &3.5 &3.7 &3.49 \cite{kehayias2013nvexp} \\
    \hline
    4H-SiC & $hh$-VV$^0$ &$\left(5\times 5 \times 2\right)$& 2.55 &2.55 &2.64 &2.72 &2.86 &3.30 \\
    &&$\left(16\times 16 \times 5\right)$&3.0 &3.0 &3.0 &3.2 &3.3 &3.30 \\
    & $kk$-VV$^0$ &$\left(5\times 5 \times 2\right)$&2.51 &2.53&2.62&2.68&2.81&2.80 \\
    &&$\left(16\times 16 \times 5\right)$&2.6 &2.6 &2.7 &2.8 &2.9 &2.80 \\
    & $hk$-VV$^0$ &$\left(5\times 5 \times 2\right)$&2.26&2.46&2.54&2.53&2.66&2.58 \\
    &&$\left(16\times 16 \times 5\right)$& 2.5&2.7 &2.8 &2.8 &2.9 &2.58 \\
  \end{tabular}
  \end{ruledtabular}
\end{table*}

We find that the HRFs computed with the PBE functional are smaller than those computed with hybrid functionals, consistent with the magnitude of the structural relaxations upon optical excitation (see Tab.~S2 of the SI). In addition, the HRFs computed with hybrid functionals are larger when phonons are obtained at the DDH (HSE) level, consistent with the fact that the phonon frequencies computed with hybrid functionals are higher than those obtained at the PBE level of theory.

\begin{figure*}
\includegraphics[width=17 cm]{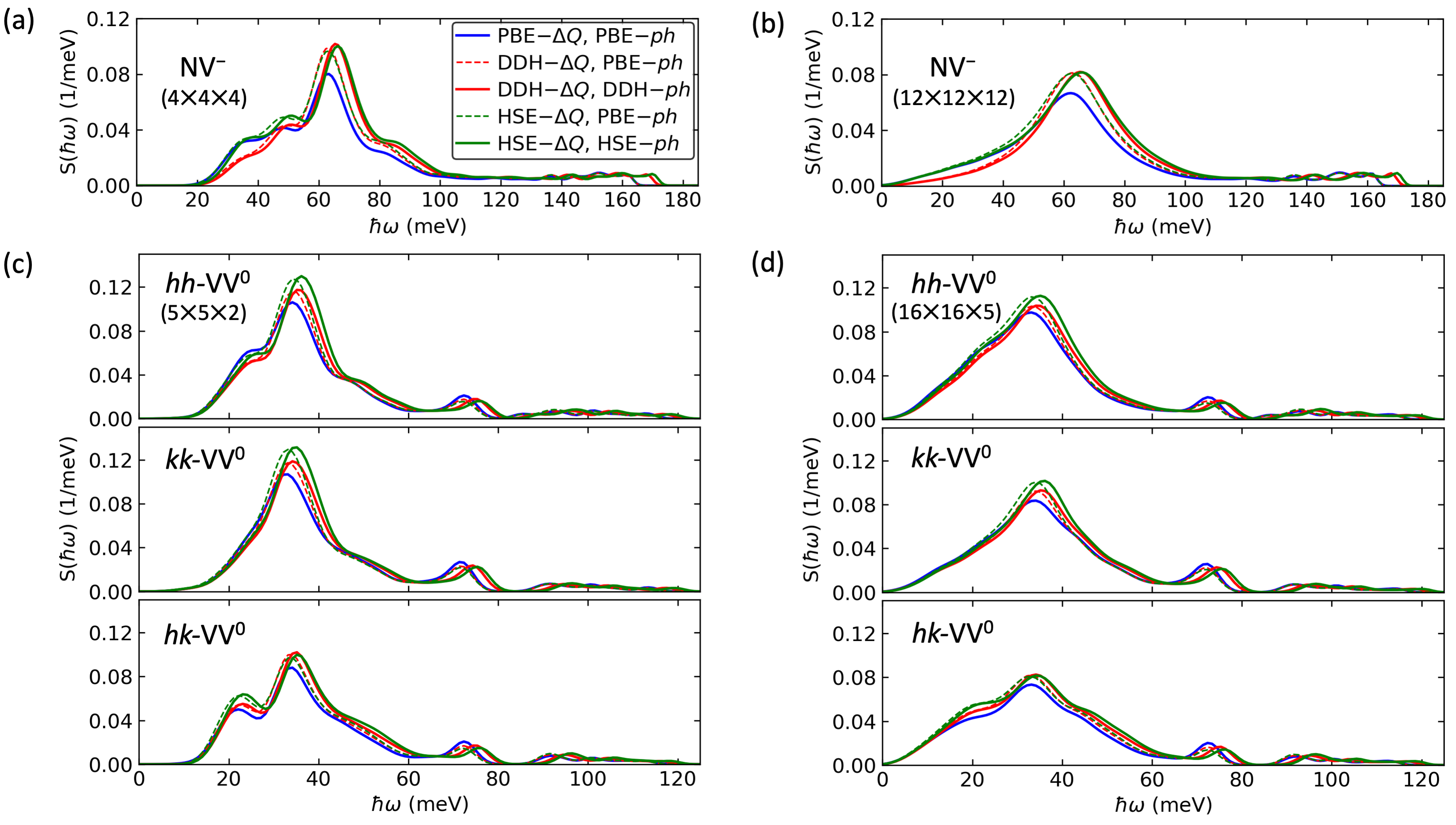}
\caption{\label{fig:HR} Spectral densities of the electron-phonon coupling, $S(\hbar\omega)$, for defect systems computed at different levels of theory. The labels follow the same notation as in Tab.~\ref{tab:hrf}. Gaussian functions with varying standard deviation ($\sigma$) were used to broaden the $\delta$-function in Eq.~\ref{eqn:e-ph-spectral}. For the NV$^-$ center (VV$^0$ centers), $\sigma$ is chosen to vary linearly from 6 (3.5) meV for the lowest-energy phonon to 1.5 (1.5) meV for the highest-energy phonon.}
\end{figure*}

The spectral densities of electron-phonon coupling (Eq.~\ref{eqn:e-ph-spectral}) are computed with the $(4 \times 4 \times 4)$ supercell for NV$^-$ and the $(5 \times 5 \times 2)$ supercell for VV$^0$ centers and shown in Fig.~\ref{fig:HR}(a) and (c). The hybrid-DFT peak intensity is higher than that computed with PBE, consistent with the values of the HRF. As for spectral densities at the level of hybrid functionals, we find that peak positions are shifted to higher energies, compared with the PBE results, when phonons are computed with hybrid functionals.

In order to evaluate finite size effects on the HRFs and spectral densities, we need to compute $\Delta Q_k$ for large supercells with either Eq.~\ref{eqn:r-dq} or Eq.~\ref{eqn:f-dq}. For the smallest supercell, Eq.~\ref{eqn:r-dq} and Eq.~\ref{eqn:f-dq} yield results that differ by less than 3\% (see Fig.~S6 of the SI); hence for larger supercells we used Eq.~\ref{eqn:f-dq} which converges more rapidly as a function of the distance from the defect center due to the fact that inter-atomic interactions in diamond and 4H-SiC are short-ranged~\cite{alkauskas2014luminescence, elisa2018siv,razinkovas2020vibrational}. Previous work on NV$^-$ has shown that forces on atoms that are separated from the defect by more than 5 \AA{} yield a negligible contribution to the HRF~\cite{razinkovas2020vibrational}. For VV$^0$ centers, we compared results for the HRF and spectral density using forces from a $(7 \times 7 \times 2)$ supercell and those from a $(5 \times 5 \times 2)$ supercell (see Fig.~S7 of the SI). We found a difference less than 5\% in the HRFs and the spectral densities are almost identical. In order to obtain phonon frequencies and modes for large supercells, we employed the force constant matrix embedding approach~\cite{alkauskas2014luminescence}, and details can be found in Sec.~V of the SI.

The HRFs computed in the dilute limit for NV$^-$ ($(12\times12\times12)$ supercell) and VV$^0$ centers ($(16\times16\times5)$ supercell) differ by 2\% and 5-15\% respectively, relative to those obtained with $(4\times4\times4)$ and $(5\times5\times2)$ supercells. Computed HRFs for NV$^-$ are in close agreement with previous theoretical predictions~\cite{alkauskas2014luminescence,gali2017nvdjt,razinkovas2020vibrational} and experiments~\cite{kehayias2013nvexp}. Computed HRFs for VV$^0$ centers are also in close agreement with our experiments. A previous theoretical work by Hashemi et al.~\cite{arsalan2021vvpl} reported a HRF of 2.75 for $hh$-VV$^0$. This value falls in the range of our results computed with the $(5\times5\times2)$ supercell and is 15\% smaller than our experimental value likely due to finite size effects not being fully taken into account in Ref.~\cite{arsalan2021vvpl}. In the dilute limit, we found that the spectral densities are smoother and exhibit a linear tail below 20 meV, reflecting the coupling with long-range acoustic phonons. Detailed analysis of spectral densities and vibrational modes can be found in Sec.~VI of the SI.

\subsection{\label{subsec:pl}Photoluminescence line shapes and Debye-Waller factors}

\begin{figure*}
\includegraphics[width=16 cm]{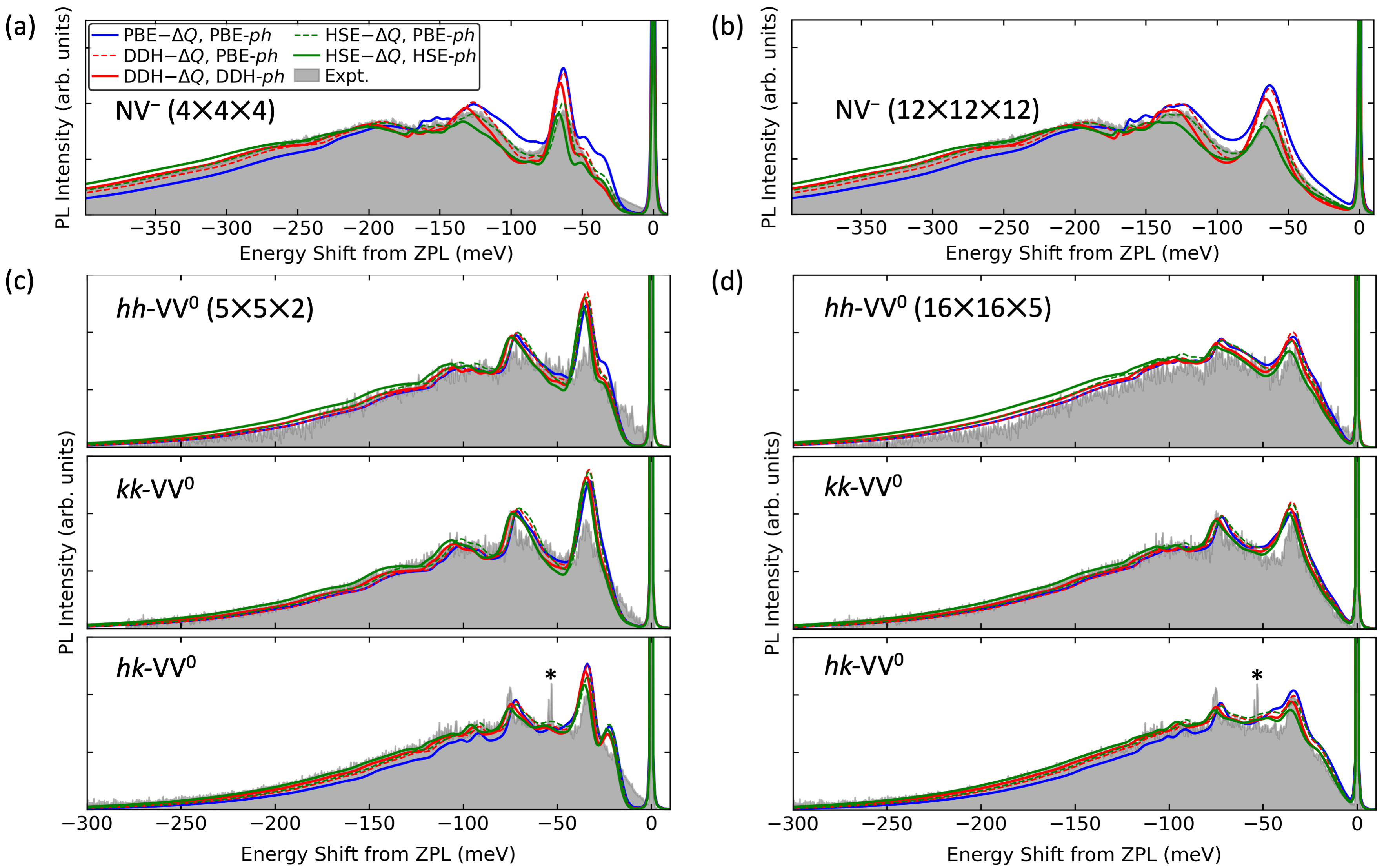}
\caption{\label{fig:PL} Photoluminescence (PL) line shapes computed at different levels of theory at low temperature (8 K for the NV$^-$ center in diamond and 10 K for VV$^0$ centers in 4H-SiC). The labels follow the same notation as in Tab.~\ref{tab:hrf}. We used $\lambda = 0.3$ (0.1) meV in Eq.~\ref{eq:genfunct} for the NV$^-$ center (VV$^0$ centers) to reproduce the experimental broadening of the zero-phonon line. The experimental data for the NV$^-$ center in diamond are from Ref.~\cite{alkauskas2014luminescence}. The small peak marked with a star `$\ast$' in the experimental curve is the ZPL of another center and should be disregarded in the comparison between theory and experiment presented here. The intensity of the experimental line shapes has been scaled to match the peaks of the computed line shapes.}
\end{figure*}

In Fig.~\ref{fig:PL} we show the PL line shapes computed using the generating function approach (Eq.~\ref{eq:Lgenfun}) with HRFs computed at different levels of theory and different supercell sizes, compared with experiment. The spectra consist of a sharp ZPL and a structured PSB. The broadening of the PSB is $\sim$500 meV for NV$^-$ and $\sim$300 meV for VV$^0$ centers. Note that PL line shapes computed with small supercells show sharper peaks and a gap of 5-10 meV between the ZPL and the PSB. The peaks located at 30 (23) meV from the ZPL for NV$^-$ ($hk$-VV$^0$) also stem from finite size effects. In our calculations, the contribution of the $e$ type phonons is computed using the HR theory, which was shown to represent an accurate approximation in the recent work by Razinkovas et al.~\cite{razinkovas2020vibrational}. These authors computed the contribution of $e$ type phonons to the PL and absorption line shapes for NV$^-$ by explicitly solving the multi-mode $E \otimes e$ Jahn-Teller problem; they showed that if the HRF of $e$ type phonons lies between 0.5 and 1.0, then the HR theory yields reasonable results for the PL line shape. For the NV$^-$ and VV$^0$ centers, we find a HRF of $e$ type phonons of about 0.5 and 1.0, respectively, indicating that the HR theory should be accurate.

We find that computed and measured line shapes agree well,  both in terms of the peak positions and intensity of line shapes, when calculations are performed with the largest supercells ($(12\times12\times12)$ for NV$^-$ and $(16\times16\times5)$ supercell for VV$^0$ centers). When computing phonons with the PBE functional, we find that the PSB is clearly dominated by coupling with the 63 meV phonon for NV$^-$, close to the experimental value 64 meV~\cite{kehayias2013nvexp}. The PSB also shows peaks at 122 meV, 135 meV, 150 meV and 161 meV  from the ZPL, in good agreement with the experimental values 122 meV, 138 meV, 153 meV and 163 meV~\cite{kehayias2013nvexp}. As for the relative intensity of peaks, the best agreement with experiment is obtained using HSE for the calculation of $\Delta Q$ and PBE phonons, respectively. When phonons are computed at the DDH (HSE) level, the 63 meV peak is shifted to 66 (67) meV, and the 161 meV peak to 168 (170) meV. Overall the agreement with experiment is good in all cases.

Similar conclusions can be drawn for the VV$^0$ centers. When computing phonons with PBE, we obtain peak positions at 34 meV and 72 meV from the ZPL in the PSB, in good agreement with experiments. Calculated PSB also exhibits small peaks located at 90 meV from the ZPL, originating from the coupling with high energy phonons. As for the relative intensity of peaks, those computed with DDH $\Delta Q$ or HSE $\Delta Q$ are in slightly better agreement with experiments than with PBE, but generalizations to other defects or materials are difficult to make. When computing phonons with the DDH or HSE functionals, the peaks are slightly shifted to lower energy by 1-3 meV, depending on the peak  and the agreement with experiments is improved, although again in all cases we find good agreement with the measured spectra. We emphasize that the ability to resolve phonon side bands is important in order to build predictive capabilities to identify fingerprints of defects using first principle calculations. 

From the PL line shape we can obtain the DWF, which is defined as the ratio of the emitted light from the ZPL to the total emitted light. At very low temperature, the DWF is computed as $\text{DWF} = e^{-S} = e^{-\sum_k S_k}$. In Tab.~\ref{tab:dwf} we report the DWFs evaluated for the largest supercells. The experimental and theoretical DWFs (see Sec.~\ref{sub-sec:expt.}) are in good agreement, and in the case of the VV$^0$ centers, we find that the computed DWFs show the trend $hh$-VV$^0$ $<$ $kk$-VV$^0$, consistent with experiments. The relation $kk$-VV$^0$ $<$ $hk$-VV$^0$ can be reproduced at the PBE level of theory. The computed DWF for $hh$-VV$^0$ ranges from 3.6\% to 5.3\%, smaller than that previously reported, 6.39\%~\cite{arsalan2021vvpl}, likely due to an incomplete finite size extrapolation in Ref.~\cite{arsalan2021vvpl}. We also note that a recent experiment on VV$^0$ centers reported DWFs of 9\% and 10\% for $kk$-VV$^0$ and $hk$-VV$^0$~\cite{shang2021vvexp}, which we consider here to be overestimates, based on our computed and measured spectra. Overall, when using hybrid functionals to compute the GS and ES electronic structure we obtain a good agreement with experiments, with small differences between results for phonons obtained with PBE or hybrid-functionals, provided an extrapolation to the dilute limit is performed.

\begin{table}
  \caption{Computed Debye-Waller factor (DWF) (\%) for spin-conserving transitions using different levels of theory. Results computed with the $(12\times12\times12)$ supercell for the NV$^-$ center in diamond and $(16\times16\times5)$ supercell for VV$^0$ centers in 4H-SiC are shown. We used the same notation as in Tab.~\ref{tab:hrf} to denote different levels of theory.}
  \label{tab:dwf}
  \begin{ruledtabular}
  \begin{tabular}{cccccc}
    \multicolumn{2}{c}{Hosts} & Diamond & \multicolumn{3}{c}{4H-SiC}  \\
    \multicolumn{2}{c}{Defects} &NV$^-$ &$hh$-VV$^0$ & $kk$-VV$^0$& $hk$-VV$^0$ \\
    \hline
    PBE$-\Delta Q$&PBE$-ph$&5.0&5.2&7.1&8.5\\
    \hline
    DDH$-\Delta Q$&PBE$-ph$&4.1&5.3&7.2&6.8\\
    &DDH$-ph$&3.7&4.8&6.6&6.1\\
    \hline
    HSE$-\Delta Q$&PBE$-ph$&3.0&4.2&6.3&6.2\\
    &HSE$-ph$&2.5&3.6&5.5&5.4\\
    \hline
    \multicolumn{2}{c}{Expt.}&3.2 \cite{alkauskas2014luminescence}&3.69&6.11&7.54
  \end{tabular}
  \end{ruledtabular}
\end{table}

\subsection{Temperature dependent photoluminescence line shapes}
\begin{figure*}
\includegraphics[width=17 cm]{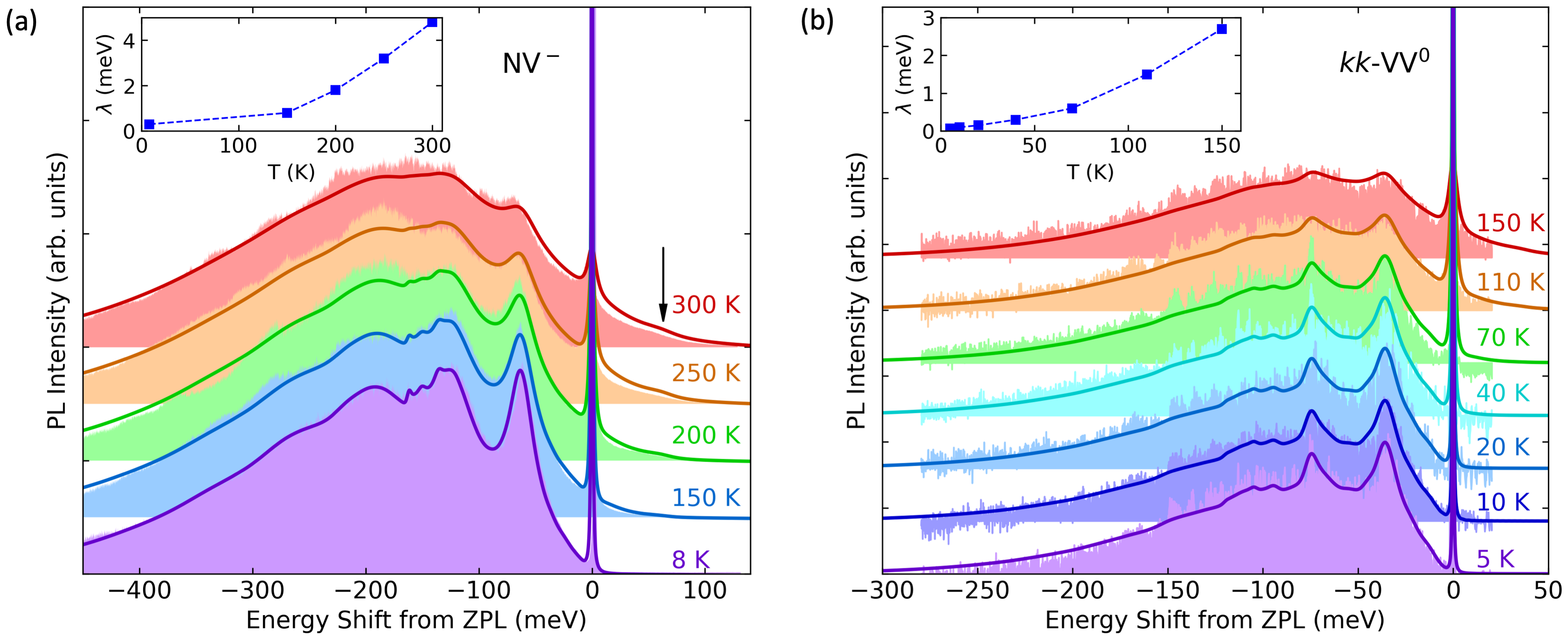}
\caption{\label{fig:TD-PL-NV} Computed photoluminescence (PL) line shapes (solid lines) of (a) the NV$^-$ center in diamond and (b) the $kk$-VV$^0$ center in 4H-SiC as a function of temperature. The best agreement with experiments is obtained when Huang-Rhys factors (HRFs) are calculated with HSE$-\Delta Q$ and PBE$-ph$ (DDH$-\Delta Q$ and DDH$-ph$) for the NV$^-$ center ($kk$-VV$^0$ center) (see Tab.~\ref{tab:hrf}). The experimental line shapes of the NV$^-$ center are averaged over twenty measurements at 8 K and 300 K and over two measurements at 150 K, 200 K and 250 K in (a). The 8 K data for the NV$^-$ center in diamond in (a) is from Ref.~\cite{alkauskas2014luminescence}. The broadening parameter $\lambda$ used in Eq.~\ref{eq:genfunct} for the theoretical line shapes is shown in the insets as a function of temperature. The black arrow in (a) indicates a shoulder at approximately 60 meV.}
\end{figure*}

In Fig.~\ref{fig:TD-PL-NV}(a) we show PL line shapes as a function of temperature for NV$^-$. We included the temperature effect on the phonon population and the line shape broadening using Eq.~\ref{eq:Lgenfun}. We tuned the parameter $\lambda$ to obtain the best agreement with experiment; this parameter describes the broadening of the line shape, and is related to the lifetime of the ES and the variation of the local environment of the defect in the experimental samples. The values of $\lambda$ obtained in our fit to experimental data are reported in the inset of Fig.~\ref{fig:TD-PL-NV}(a). We approximated $\lambda$ with a quadratic function of $T$ for NV$^-$, consistent with the findings of Ref.~\cite{fukami2019nvthermomentry} for $T\geq 100$ K. We note that the ZPL width was shown to depend on $T^5$ due to the dynamic Jahn-Teller effect of the $^3E$ ES for $T\leq 80$ K~\cite{fu2009nvdjt,abtew2011nvdjt}. The $T^5$ temperature dependence was not considered in our work due to the lack of experimental data.

Overall, the calculated temperature-dependent PL line shapes agree well with experiments. As the temperature increases, the ZPL width increases due to the decrease of the lifetime of the ES. The intensity of the PSB in the 30 meV range around the ZPL also increases, indicating an increasing population of the higher vibrational levels of the ES long-range modes. The increasing population causes the broadening and the small shift of the first peak of the PSB (about 63 meV lower than ZPL) towards lower energies, as observed both theoretically and experimentally. As for the PSB with energy higher than the ZPL, we find that a shoulder peak at about 60 meV becomes increasingly more intense as the temperature increases, due to the coupling with the quasi-local mode in the ES. Temperature effects on the electronic structure, atomic structure and lattice parameters were neglected in our calculations. These effects are assumed to be relatively small considering the $\sim$3 meV shift of the ZPL from 8 K to 300 K observed in experiments.

The measured and computed temperature-dependent PL line shapes for $kk$-VV$^0$ (Fig.~\ref{fig:TD-PL-NV}(b)) are in general good agreement. Also in this case, we observe a broadening of the ZPL and the PSB and the increase of the intensity around the ZPL as the temperature increases. The chosen broadening parameter $\lambda$ turns out to be a non-linear function of the temperature. Our temperature dependent results show that converged calculations can successfully discern features in the PSB also as a function of $T$.

\subsection{Displaced harmonic oscillator and the Franck-Condon approximations}

We have presented results for PL line shapes obtained using the generating function approach (Eq.~\ref{eq:genfunct}), which in turn was derived using the FC and the DHO approximations. The former assumes that the transition dipole moment $|\boldsymbol{\mu}_{eg}|$ is independent of changes in the atomic structure, and the latter assumes that the vibrational modes of the GS and the ES are identical except for a displacement. We present below an analysis of the validity of these two approximations using a one-dimensional (1D) model, where just one effective phonon mode is considered. Previous studies have shown that the 1D model provides an accurate description of defect systems with strong electron-phonon coupling ($\text{HRF} \gg 1$) and serves as a valuable approximation to cases with weak or intermediate electron-phonon coupling~\cite{alkauskas2012plgan,alkauskas2016tutorial}. The systems considered in this work, e.g., NV$^-$ and $kk$-VV$^0$, yield $\text{HRF}\approx 3$ (intermediate electron-phonon coupling); hence the use of the 1D model appears to be justified, although the Herzberg-Teller (HT) effect of symmetry forbidden vibrational modes and the Duschinsky rotation effect between vibrational modes of the GS and the ES~\cite{duschinsky1937} are not captured by the model.

In the 1D model, we considered one effective phonon mode which includes only vibrations projected along the direction of the configuration coordinate $Q$, which connects the equilibrium atomic structures of the GS and the ES (see Fig.~\ref{fig:schematic-pl}). The frequency of such effective phonon mode is calculated as the weighted average over all phonon frequencies in either the GS or the ES:
\begin{equation}
    \Omega_{\{e,g\}} = \dfrac{\sum_k \omega_{\{e,g\};k}^2 \Delta Q_k^2}{\sum_k \Delta Q_k^2}.
    \label{eq:Omegaeg}
\end{equation}
At the PBE level of theory we obtain:
for NV$^-$, $\Omega_g = 63.06$ meV,  $\Omega_e = 66.38$ meV, and $\Delta Q =\sqrt{\sum_k \Delta Q_k^2} = 0.653$ amu$^{0.5}$~\AA{}; for $kk$-VV$^0$, $\Omega_g = 38.11$ meV,  $\Omega_e = 43.45$ meV, and $\Delta Q = 0.763$ amu$^{0.5}$~\AA{}. With these parameters we can also compute the HRF for the 1D model as $\text{HRF}=\frac{\Omega_g \Delta Q^2}{2\hbar}$. We obtained 3.22 (2.65) for NV$^-$ ($kk$-VV$^0$), close to the HRF from the all-phonon calculation with the $(4\times4\times4)$ ($(5\times5\times2)$) supercell, which is 2.94 (2.51).

\begin{figure*}
    \centering
    \includegraphics[width=17cm]{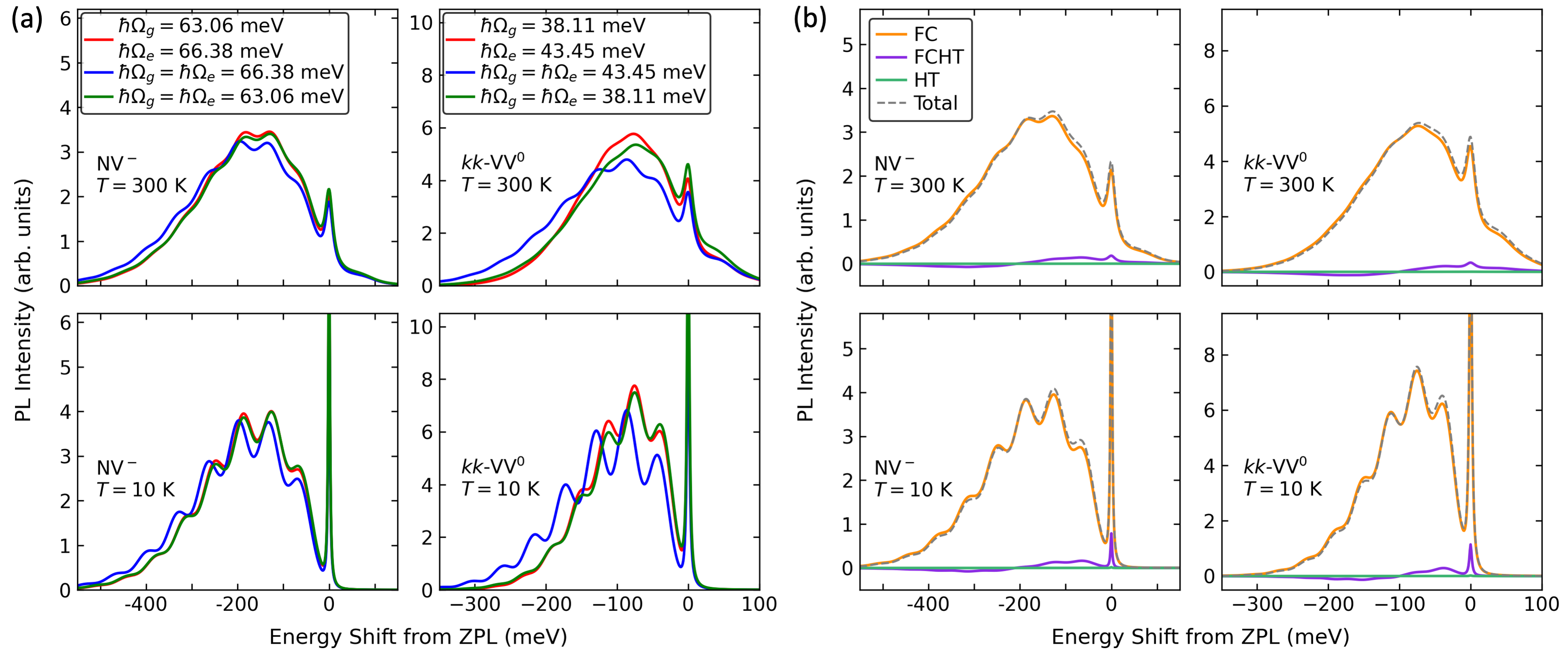}
    \caption{PL line shapes evaluated within the 1D model for the NV$^-$ center in diamond and the $kk$-VV$^0$ center in 4H-SiC at $T=10$ K and $T=300$ K. (a) Analysis of the the displaced harmonic oscillator (DHO) approximation. Red lines denote the line shapes computed with both the ground state (GS) and the excited state (ES) frequency. Blue (green) lines denote the line shapes computed with DHO approximation using the ES (GS) frequency.
    (b) `FC', `FCHT', and `HT' denote the contribution to the PL given by the Franck-Condon, Franck-Condon Herzberg-Teller, and the Herzberg-Teller terms, respectively. `Total' denotes the sum of three terms. The zero-phonon line is broadened using a Lorentzian function with scale parameter $\lambda$, and the phonon sideband is broadened using a Gaussian function with standard deviation $\sigma$. For the NV$^-$ center, $\lambda=2$ (10) meV and $\sigma=25$ (30) meV were used at $T=10$ (300) K. For the $kk$-VV$^0$ center, $\lambda=2$ (8) meV and $\sigma=15$ (20) meV were used at $T=10$ (300) K.}
    \label{fig:1DPL}
\end{figure*}

Fig.~\ref{fig:1DPL}(a) shows the PL spectra of NV$^-$ and $kk$-VV$^0$ computed using the 1D model with (i)  actual $\Omega_g$ different from $\Omega_e$, (ii) $\Omega_g$ the same as $\Omega_e$ and (iii) $\Omega_e$ the same as $\Omega_g$. We find that for the NV$^-$ center, the approximation (iii) yields 2.3\% and 3.1\% relative error at $T = 10$ K and at $T = 300$ K relative to (i). This result agrees with those reported by Razinkovas et al.~\cite{razinkovas2020vibrational}. In the case of the $kk$-VV$^0$ center, the approximation (iii) yields a 6.0\% and 9.5\% relative error at $T = 10$ K and at $T = 300$ K, respectively, arising from the greater difference between $\Omega_g$ and $\Omega_e$. The error caused by using (iii) increases as a function of $T$, since the number of excited vibrational states contributing to the PL spectrum increases. Since  excited vibrational states are more likely to be populated in $kk$-VV$^0$ than in NV$^-$, due to the smaller effective phonon frequency, the error is larger for the former. The DHO approximation used in the calculations of PL line shapes in Sec.~\ref{subsec:pl} corresponds to case (iii). Our analysis with the 1D model suggests that the DHO approximation used in our calculations is fairly accurate. In general, we expect the DHO approximation to be valid for defects in rigid materials with relatively small structural displacements upon optical transitions at low temperature.

In order to use the 1D model to examine the validity of the FC approximation, we first write Eq.~\ref{eq:Ietog} without using  the FC principle: 
\begin{equation}
\begin{aligned}
    \label{eq:1DPL-BO}
    L(\hbar\omega,T) &\propto \omega^3 \sum_i\sum_j P_{ej}(T) |\langle \Theta_{ej}|\boldsymbol{\mu}_{eg}|\Theta_{gi}\rangle|^2 \\
    &\times \delta\left( E_{\text{ZPL}} + E_{ej} - E_{gi} - \hbar\omega \right).
\end{aligned}
\end{equation}
Within the 1D effective phonon approximation we have:
\begin{equation}
\label{eq:newTheta}
     \langle \Theta_{ej}|\boldsymbol{\mu}_{eg}|\Theta_{gi}\rangle = \int dQ  \phi^\ast_{n^{ej}}(Q)\boldsymbol{\mu}_{eg}(Q)\phi_{n^{gi}}(Q).
\end{equation}
Here we use the same notations for nuclear wavefunctions and vibrational states as in Eq.~\ref{eq:nuclear-wavefunctions}. The quantity $\boldsymbol{\mu}_{eg}(Q) = \langle \psi_e(Q) | \hat{\boldsymbol{\mu}} | \psi_g(Q) \rangle$ is the transition dipole moment between electronic wavefunctions obtained at fixed values of the configuration coordinate $Q$, where $\psi_e\left(\psi_g\right)$ is the electronic wavefunction of the system in the ES (GS). To first order in $Q$ we can further approximate Eq.~\ref{eq:newTheta} as:
\begin{equation}
\begin{aligned}
\label{eq:newThetaTaylorExp}
     &\langle \Theta_{ej}|\boldsymbol{\mu}_{eg}|\Theta_{gi}\rangle \\
     &\quad\approx \boldsymbol{\mu}_{eg}(Q=0) \langle \phi_{n^{ej}}|\phi_{n^{gi}}\rangle + \left.\frac{d\boldsymbol{\mu}_{eg}}{dQ}\right|_{Q=0} \langle \phi_{n^{ej}}|Q|\phi_{n^{gi}}\rangle
\end{aligned}
\end{equation}
We have numerically computed the electronic transition dipole moment as a function of $Q$, and  verified that the dependence is linear with a relative change of about 10\% between the equilibrium atomic structures of GS and ES for both NV$^-$ and $kk$-VV$^0$ (see Fig.~S12 of the SI).

After introducing Eq.~\ref{eq:newThetaTaylorExp} in  Eq.~\ref{eq:1DPL-BO} we recognize the usual FC term, and the terms beyond the FC approximation. The latter can be grouped into two categories: the Franck-Condon Herzberg-Teller (FCHT), and the Herzberg-Teller (HT) term, depending on whether one or two derivatives of the electronic transition dipole moment with respect to $Q$ are present, respectively (see Eq.~S13-S15 of the SI). Fig.~\ref{fig:1DPL}(b) shows the FC, FCHT, and HT contributions to the PL line shape for both NV$^-$ and $kk$-VV$^0$, at $T = 10$ K and $T = 300$ K. We find that the FC term is the dominant one and that the FCHT and HT contributions are smaller than 5\% and 0.1\% of the total intensity, respectively. These results indicate that the FC approximation used in Sec.~\ref{subsec:pl} is accurate. We note that the validity of the FC approximation depends on the symmetry and the strength of electron-phonon coupling of the defect center. For negatively-charged silicon vacancy centers in diamond, the HT term may not be negligible~\cite{elisa2018siv,gali2018groupiv}. We suggest that computing the relative error caused by neglecting FCHT and HT contributions using the 1D model is a useful first step in assessing the validity of the FC approximation. 

\section{\label{sec:conclusion}Conclusions}

In summary, we presented a detailed comparison of measured and computed PL spectra of defects in diamond and SiC, aimed at assessing the validity of theoretical and numerical approximations used in first principles calculations. As expected, our results show that the best agreement between theory and experiments is obtained when using hybrid functionals, instead of PBE, although the qualitative differences between the PL line shapes of the different configurations of the VV$^0$ centers are reproduced with PBE as well. We find minor differences between the results obtained with the hybrid functionals HSE and DDH: the values of the ZPL obtained with the two functionals are almost identical (note that our HSE results slightly differ from previous ones reported in the literature for the case of the NV$^-$) and the values of the HRF differ by less than 10\%. We also find minor differences between spectral densities computed at the PBE and DDH level of theory, indicating that the major improvement of hybrid functionals over PBE is in the determination of the electronic structure of the system. Our findings show that results for the triplet ES obtained with CDFT and TDDFT are similar at the DDH level of theory for NV$^-$ and VV$^0$ centers, suggesting that the relatively cheap CDFT method is accurate for the calculations of their PL spectra. In addition, by using a 1D model, we provided a qualitative assessment of the approximation arising from the use of the FC principle and of the DHO approximation, finding that both of them are justified. Finally, we emphasize the importance of finite size scaling to obtain theoretical results in agreement with experiments for HRFs and PL line shapes, especially for the contribution of quasi-local and long-range acoustic phonon modes. The protocol established in our work shows that accurate results for PL spectra may be obtained at a given temperature using the generating function approach, with phonons extrapolated to the dilute limit, and by using hybrid functionals to compute the GS and ES potential energy surfaces of the defects, with the ES computed with constrained DFT. A 1D model can be used to evaluate the accuracy of the FC and DHO approximations. This protocol, validated here for NV$^-$ and VV$^0$ centers, leads to robust predictions of the overall line shape, including PSBs, which can be used to aid the identification and characterization of optically-active defects.

\begin{acknowledgments}
We thank He Ma and Masaya Fukami for helpful discussion, and Bob B. Buckley for assistance with the initial diamond NV$^-$ center measurements. Theoretical and computational work (Y.J., M.G., G.G.) and experimental validation at Argonne (S.E.S., F.J.H.) was supported by the Midwest Integrated Center for Computational Materials (MICCoM) as part of the Computational Materials Sciences Program funded by the U.S. Department of Energy. Additional experimental efforts (G.W., D.D.A.) were supported by the U.S. Department of Energy, Office of Science, Basic Energy Sciences, Materials Sciences and Engineering Division. This research used resources of the National Energy Research Scientific Computing Center (NERSC), a DOE Office of Science User Facility supported by the Office of Science of the U.S. Department of Energy under Contract No. DE-AC02-05CH11231, resources of the Argonne Leadership Computing Facility, which is a DOE Office of Science User Facility supported under Contract No. DE-AC02-06CH11357, and resources of the University of Chicago Research Computing Center. 

\end{acknowledgments}


\bibliographystyle{apsrev4-2}
\bibliography{main}

\end{document}


\preprint{APS}

\title{Supplementary Information\\ for the paper entitled\\ Photoluminescence spectra of point defects in semiconductors: validation of first principles calculations
}

\author{Yu Jin}
\affiliation{Department of Chemistry, University of Chicago, Chicago, Illinois 60637, United States}
\author{Marco Govoni}
\email{mgovoni@anl.gov}
\affiliation{Pritzker School of Molecular Engineering, University of Chicago, Chicago, Illinois 60637, United States}
\affiliation{Materials Science Division and Center for Molecular Engineering, Argonne National Laboratory, Lemont, Illinois 60439, United States}
\author{Gary Wolfowicz}
\affiliation{Materials Science Division and Center for Molecular Engineering, Argonne National Laboratory, Lemont, Illinois 60439, United States}
\author{Sean E. Sullivan}
\affiliation{Materials Science Division and Center for Molecular Engineering, Argonne National Laboratory, Lemont, Illinois 60439, United States} 
\author{\\ F. Joseph Heremans}
\affiliation{Pritzker School of Molecular Engineering, University of Chicago, Chicago, Illinois 60637, United States}
\affiliation{Materials Science Division and Center for Molecular Engineering, Argonne National Laboratory, Lemont, Illinois 60439, United States}
\author{David D. Awschalom}
\affiliation{Pritzker School of Molecular Engineering, University of Chicago, Chicago, Illinois 60637, United States}
\affiliation{Materials Science Division and Center for Molecular Engineering, Argonne National Laboratory, Lemont, Illinois 60439, United States}
\affiliation{Department of Physics, University of Chicago, Chicago, Illinois 60637, United States}
\author{Giulia Galli}
\email{gagalli@uchicago.edu}
\affiliation{Department of Chemistry, University of Chicago, Chicago, Illinois 60637, United States}
\affiliation{Pritzker School of Molecular Engineering, University of Chicago, Chicago, Illinois 60637, United States}
\affiliation{Materials Science Division and Center for Molecular Engineering, Argonne National Laboratory, Lemont, Illinois 60439, United States}

\date{\today}

\maketitle

\section{\label{s-sec:bulk} Bulk diamond and 4H-SiC}
Bulk properties including lattice constants and band gaps are computed for diamond and 4H-SiC at different levels of theory and summarized in Tab.~\ref{s-tab:bulk}. Band gaps computed at the level of DDH or HSE are in good agreement with experiments. 
The best agreement with the experimental  lattice constants is obtained using the PBE and the HSE functional for diamond and 4H-SiC, respectively. We observe that bonds predicted by HSE are stiffer than the ones obtained with DDH, which are in turn stiffer than PBE ones.

\begin{table}[b]
\caption{\label{s-tab:bulk}%
Bulk properties of diamond and 4H-SiC computed using PBE, DDH and HSE functionals. Experimental values are also shown.
}
\begin{ruledtabular}
\begin{tabular}{ccccc}
Diamond&PBE&DDH&HSE&Expt.\\
\colrule
$a$ (Å)&3.568&3.55\footnote{Ref.~\cite{ddh_PRB_2014}}&3.543\footnote{Ref.~\cite{hseo_PRM_2017}}&3.567\footnote{Ref.~\cite{diamond_exp_lp}}\\
$E_g$ (eV) &4.19&5.59&5.42&5.48\footnote{Ref.~\cite{diamond_exp_gap}}\\
\colrule
4H-SiC&PBE&DDH$^{\text{b}}$&HSE$^{\text{b}}$&Expt.\footnote{Ref.~\cite{sic_gap_lp}}\\
\colrule
$a$ (Å)&3.095&3.087&3.074&3.073\\
$c$ (Å)&10.133&10.089&10.074&10.053\\
$E_g$ (Å)&2.27&3.28&3.19&3.23\\
\end{tabular}
\end{ruledtabular}
\end{table}

\section{\label{s-sec:cdft} Details of CDFT calculations}

\begin{figure*}
\includegraphics[width=16 cm]{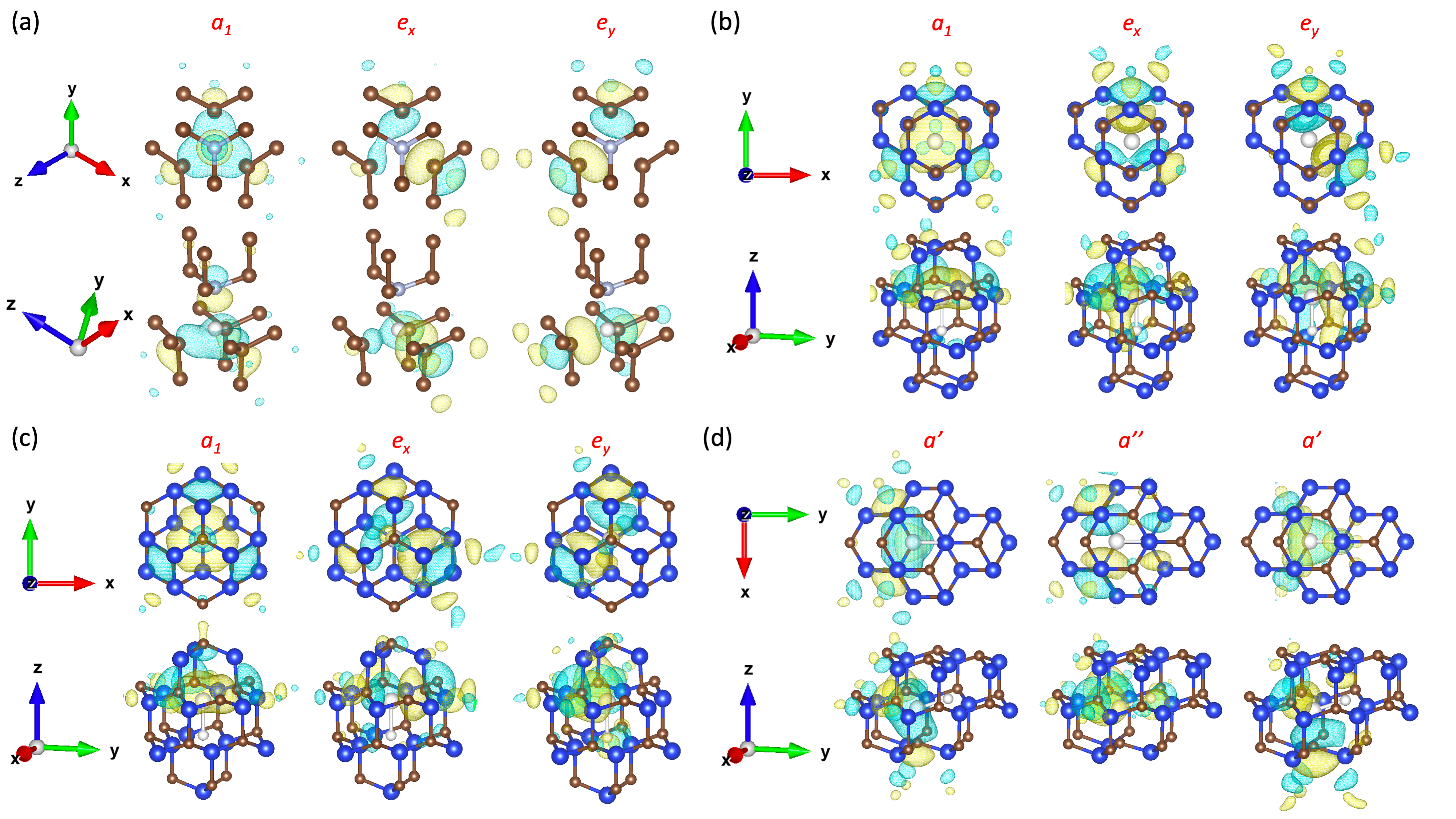}
\caption{\label{s-fig:isosurface-orbitals} Isosurfaces of the square of module of Kohn-Sham orbitals associated to defect states for (a) NV$^-$ in diamond, (b) $hh$-VV$^0$ in 4H-SiC, (c) $kk$-VV$^0$ in 4H-SiC, and (d) $hk$-VV$^0$ in 4H-SiC. The isosurface level is set to 0.015 e/\AA$^3$. The color (yellow or light below) represents the sign $(+/-)$ of the orbital. Blue, brown, silver, and white spheres denote Silicon, Carbon, Nitrogen atoms and vacancies. The $+x$, $+y$ and $+z$ axes are shown using the compass for both the top view and the side view.}
\end{figure*}

The Kohn-Sham orbitals of defect states computed with DFT using the DDH functional are reported in Fig.~\ref{s-fig:isosurface-orbitals}. For the NV$^-$ in diamond the $a_1$ orbital and $e$ orbitals are mainly localized on the three carbon atoms that are first neighbors of the carbon vacancy ($\text{V}_\text{C}$). For VV$^0$ centers in 4H-SiC, the $a_1$ orbital and the lower-energy $e$ orbitals are mainly localized on the three carbon atoms that are first neighbors of the silicon vacancy ($\text{V}_\text{Si}$).

\begin{figure*}
\includegraphics[width=16 cm]{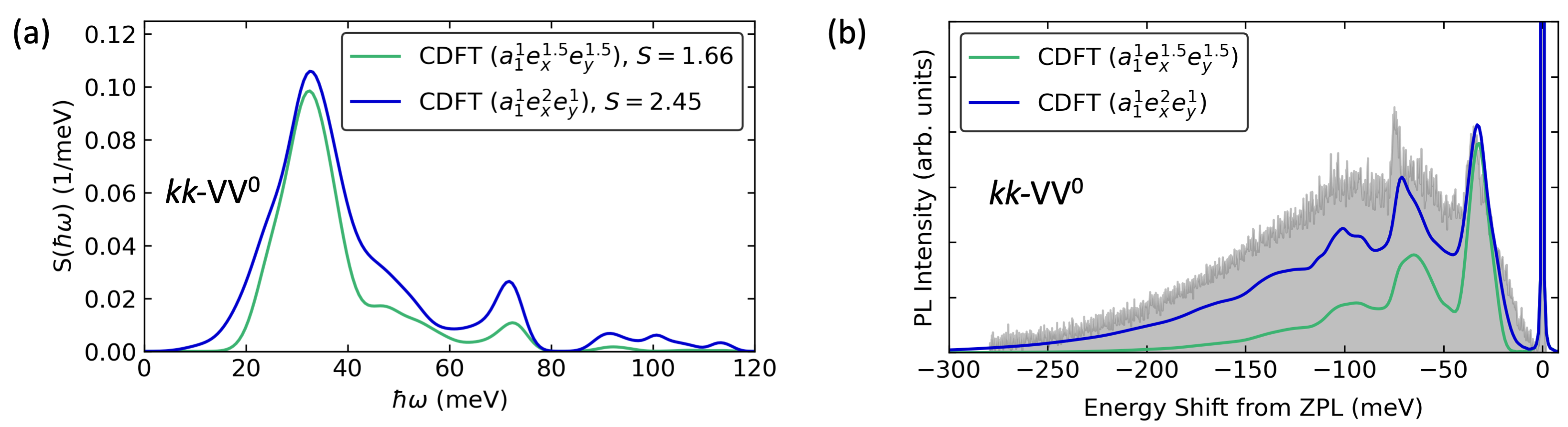}
\caption{\label{s-fig:cdft-occupations} Influence of the occupations in CDFT excited-state (ES) calculations on the results. (a) Computed spectral densities of electron-phonon coupling  $S(\hbar\omega)$ for the $kk$-VV$^0$ center in 4H-SiC. The total Huang-Rhys factor (HRF, $S$) is given in the legend. (b) Computed photoluminescence (PL) line shapes for the $kk$-VV$^0$ center in 4H-SiC. The gray area represents the experimental PL line shape measured at 10 K. Calculations were performed with $(5\times5\times2)$ supercell at the PBE level of theory.}
\end{figure*}

At the PBE level of theory, we examined several ways for constraining occupations in CDFT calculations in order to accurately represent the optical transition from the $a_1$ to the degenerate $e$ orbitals in the $kk$-VV$^0$ in 4H-SiC. The triplet ground state (GS) can be denoted as configuration $a_1^2e_x^1e_y^1$. After applying a spin conserving excitation, the triplet excited state (ES) is in the configuration $a_1^1e_x^2e_y^1$ (equivalent to  $a_1^1e_x^1e_y^2$). When the latter configuration is constrained within CDFT, the system ES is forced to have $C_{1h}$ symmetry, which corresponds to the actual local minimum on the adiabatic ES potential energy surface. A commonly adopted strategy to enforce $C_{3v}$ symmetry, and therefore simulate an ES whose atomic structure relaxation is not coupled to $e$ type phonons, is to constrain CDFT with the configuration $a_1^1e_x^{1.5}e_y^{1.5}$. The energy of the ZPL ($E_{\text{ZPL}}$) and the atomic displacement computed for the ES with $C_{3v}$ symmetry are 0.04 eV (4\%) greater and 18\% smaller than the ones computed for the ES with $C_{1h}$ symmetry. As a result, the computed HRFs, the spectral densities of electron-phonon coupling, and PL line shapes are significantly different depending on the chosen constrained occupations, as shown in Fig.~\ref{s-fig:cdft-occupations}. To correctly describe the coupling with $e$ type phonons in the optical process, $a_1^1e_x^2e_y^1$ occupations should be used for CDFT calculations.

At the DDH/HSE level of theory, we observed that the self-consistent cycle within CDFT could not converge by using the configuration $a_1^1e_x^2e_y^1$. In such cases, we performed CDFT calculations with the configuration $a_1^1e_x^{1.5}e_y^{1.5}$, which did not show convergence issues. We then added a correction term to determine the energy of the ZPL ($E_{\text{ZPL}}$) and the equilibrium structure of the ES based on the results obtained at the PBE level of theory:
\begin{equation}
\begin{aligned}
    E_{\text{ZPL}}\left(a_1^1e_x^2e_y^1, \mathrm{DDH/HSE}\right) & =
    E_{\text{ZPL}}\left(a_1^1e_x^{1.5}e_y^{1.5}, \mathrm{DDH/HSE}\right)\\
    &+ \left[ E_{\text{ZPL}}\left(a_1^1e_x^{2}e_y^{1}, \mathrm{PBE}\right) - E_{\text{ZPL}}\left(a_1^1e_x^{1.5}e_y^{1.5}, \mathrm{PBE}\right)\right], \\
    \mathbf{R} \left(a_1^1e_x^2e_y^1, \mathrm{DDH/HSE}\right) &= \mathbf{R}\left(a_1^1e_x^{1.5}e_y^{1.5}, \mathrm{DDH/HSE}\right)\\
    &+ \left[ \mathbf{R}\left(a_1^1e_x^{2}e_y^{1}, \mathrm{PBE}\right) - \mathbf{R}\left(a_1^1e_x^{1.5}e_y^{1.5}, \mathrm{PBE}\right)\right],
\end{aligned}
\end{equation}
where $E_{\text{ZPL}}\left(a_1^1e_x^{1.5}e_y^{1.5}, \mathrm{PBE/DDH/HSE}\right)$ and $\mathbf{R}\left(a_1^1e_x^{1.5}e_y^{1.5}, \mathrm{PBE/DDH/HSE}\right)$ denote the $E_{\text{ZPL}}$ and the equilibrium structure of the ES obtained with the configuration $a_1^1e_x^{1.5}e_y^{1.5}$ at the PBE/DDH/HSE level.

The comparison between TDDFT and CDFT results with the DDH functional was carried out with the configuration $a_1^1e_x^{1.5}e_y^{1.5}$ in CDFT also because of the convergence issues with the configuration $a_1^1e_x^{2}e_y^{1}$. We checked that the energy difference between the CDFT calculations with different occupations at the PBE level is small ($\sim$0.04 eV). Hence we concluded that our TDDFT/CDFT comparison with DDH is meaningful and accurate to $\sim$0.04 eV.

\begin{figure}
\includegraphics[width=8 cm]{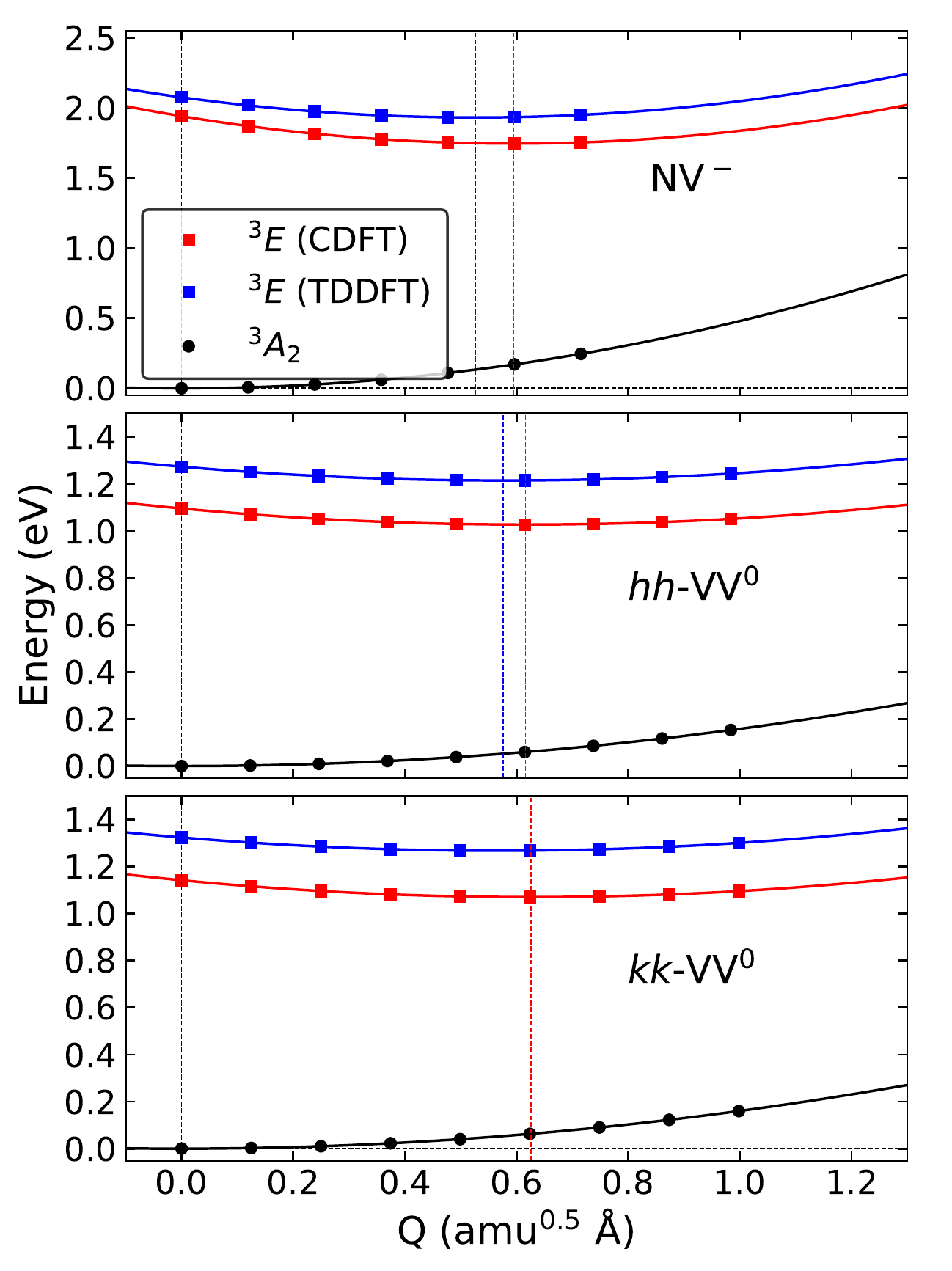}
\caption{\label{s-fig:tddft-pbe}Configuration coordinate diagrams describing the total energies of the $^3A_2$ ground state (GS) and the $^3E$ excited state (ES) along the relaxation path resulting from CDFT with the electronic configuration $a_1^1e_x^{1.5}e_y^{1.5}$ for the NV$^-$ center in diamond and the $hh$-VV$^0$ and the $kk$-VV$^0$ centers in 4H-SiC. Calculations are performed at the PBE level of theory. Dashed vertical lines denote the locations of the local minimum by fitting the energy curves with quadratic functions.}
\end{figure}

\section{\label{s-sec:displace} Atomic displacements upon optical excitation}
Total atomic displacement $\Delta R$ and mass-weighted atomic displacement $\Delta Q$ are computed for all defect systems at different levels of theory and summarized in Tab.~\ref{s-tab:displacements}.
\begin{equation}
   \begin{aligned}
   \Delta R = \left( \sum_{\alpha=1}^{N}\sum_{i=x,y,z} \Delta \mathbf{R}_{\alpha i}^2 \right)^{1/2}, \quad \Delta Q = \left( \sum_{\alpha=1}^{N}\sum_{i=x,y,z} M_{\alpha} \Delta \mathbf{R}_{\alpha i}^2 \right)^{1/2}. \\
   \end{aligned}
   \label{eqn:displacement}
\end{equation}
Here $\Delta \mathbf{R}_{\alpha i} = (\mathbf{R}_{\alpha i})_e - (\mathbf{R}_{\alpha i})_g$ is the atomic displacement of the $\alpha$-th atom in the $i$-th direction between the equilibrium structures of the ES and the GS. $M_{\alpha}$ is the mass of the $\alpha$-th atom.

\begin{table}[b]
  \caption{\label{s-tab:displacements} Displacements $\Delta R$ (\AA) and mass-weighted displacements $\Delta Q$ (amu$^{0.5}$ \AA) between the equilibrium structures of the ground state (GS) and the excited state (ES) computed at different levels of theory.}
  \label{tab:all-displacements}
  \begin{ruledtabular}
    \begin{tabular}{cccccccc}
    Hosts  & Defects & \multicolumn{2}{c}{PBE} & \multicolumn{2}{c}{DDH} & \multicolumn{2}{c}{HSE}  \\
    & & $\Delta R$ & $\Delta Q$& $\Delta R$ & $\Delta Q$ & $\Delta R$ & $\Delta Q$ \\
    \colrule
    Diamond  & NV$^{-}$ &0.187 &0.653 &0.191 &0.666 &0.200 & 0.697  \\
    \colrule
    4H-SiC & $hh$-VV$^0$ &0.186 &0.785 &0.195 &0.816 &0.200 &0.834 \\
      & $kk$-VV$^0$ &0.185 &0.763 &0.192 &0.787 &0.198 &0.813 \\
      & $hk$-VV$^0$ &0.183 &0.759 &0.190 &0.785 &0.200 &0.835 \\
  \end{tabular}
  \end{ruledtabular}
\end{table}

One can notice that the magnitude of $\Delta R$ and $\Delta Q$ follows the relation: HSE $>$ DDH $>$ PBE, which can be related to the stiffness of the bonds. For VV$^0$ centers in 4H-SiC, $\Delta R$ and $\Delta Q$ follow the relation: $hh$-VV$^0$ $>$ $kk$-VV$^0$ $>$ $hk$-VV$^0$ in most cases. The same relation is followed by the HRFs, as discussed in the main text Sec.~IV~B.

\begin{figure}[b]
\includegraphics[width= 8 cm]{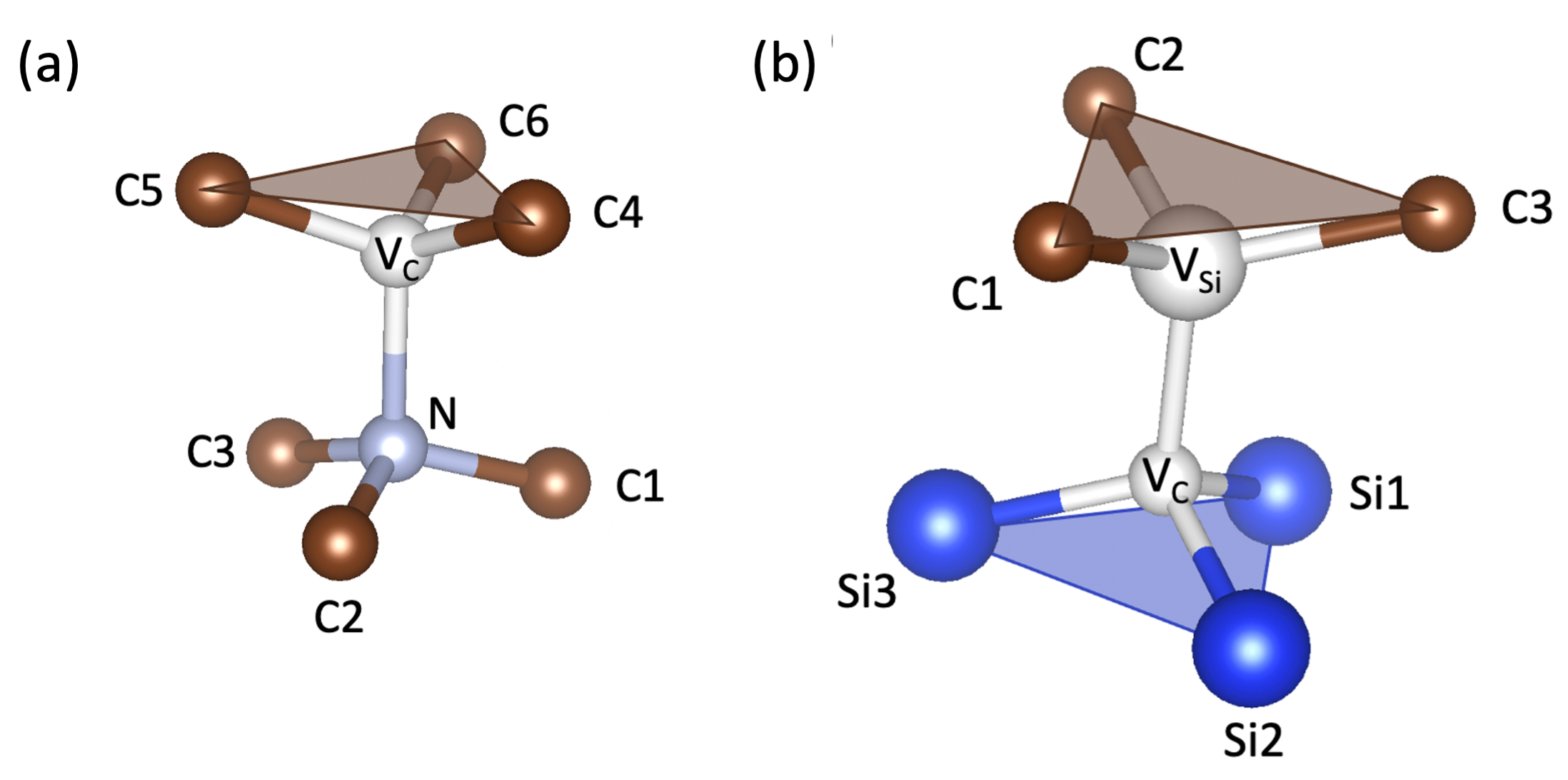}
\caption{\label{s-fig:detailed-structure} Atomic structure of (a) the NV$^-$ center in diamond and (b) the $kk$-VV$^0$ center in 4H-SiC. Carbon, nitrogen and silicon atoms are represented using brown, sliver and blue spheres. Vacancy sites are represented using white spheres. Labels of atoms are used in Tab.~\ref{s-tab:detailed-nv-dis} and Tab.~\ref{s-tab:detailed-kk-vv-dis}.}
\end{figure}

We examined the atomic displacements of neighbor atoms upon optical excitation for the NV$^-$ center in diamond and the $kk$-VV$^0$ center in 4H-SiC. For the NV$^-$ center, the distances between three carbon atoms around the $\mathrm{V}_{\text{C}}$ were computed for both the GS and the ES and summarized in Tab.~\ref{s-tab:detailed-nv-dis}. In the GS structure, three pairs of carbon atoms around the $\mathrm{V}_{\text{C}}$ have the same distances due to the $C_{3v}$ symmetry. In the ES structure, all distances and bond lengths are increased because the $e$ defect orbitals are more delocalized that $a_1$ orbitals. One pair of carbon atoms has longer distances while the other two pairs have shorter distances, indicating that the symmetry is reduced from $C_{3v}$ to $C_{1h}$. Length of three nitrogen-carbon bonds also show a similar behavior, but the magnitude of the asymmetric stretching is much smaller. This is consistent with the fact that the defect orbitals are mainly localized on the carbon atoms around the $\mathrm{V}_{\text{C}}$.

\begin{table}
  \caption{Distances (\AA) between neighbor atoms around the NV$^-$ center in diamond in the equilibrium structures of the ground state (GS) and the excited state (ES). N is the nitrogen substituent, and C1, C2 and C3 are the three carbon atoms connected to it. C4, C5 and C6 are three carbon atoms adjacent to the carbon vacancy site, as shown in Fig.~\ref{s-fig:detailed-structure}.}
  \label{s-tab:detailed-nv-dis}
  \begin{ruledtabular}
  \begin{tabular}{ccccccc}
    Atom pairs & \multicolumn{2}{c}{PBE} & \multicolumn{2}{c}{DDH} & \multicolumn{2}{c}{HSE}  \\
    & GS & ES & GS & ES & GS & ES \\
    \colrule
    N$-$C1 &1.472 &1.491 &1.463 &1.482 &1.460 &1.479 \\
    N$-$C2 &1.472 &1.490 &1.463 &1.481 &1.460 &1.478 \\
    N$-$C3 &1.472 &1.490 &1.463 &1.481 &1.460 &1.478 \\
    C4$-$C5 &2.664 &2.762 &2.639 &2.749 &2.631 &2.744 \\
    C4$-$C6 &2.664 &2.808 &2.639 &2.795 &2.631 &2.790 \\
    C5$-$C6 &2.664 &2.762 &2.639 &2.749 &2.631 &2.744 \\
  \end{tabular}
  \end{ruledtabular}
\end{table}

For the $kk$-VV$^0$ center, the distance between three carbon atoms around the $\mathrm{V}_{\text{Si}}$ and the distance between three silicon atoms around the $\mathrm{V}_{\text{C}}$ were computed for both the GS and the ES and summarized in Tab.~\ref{s-tab:detailed-kk-vv-dis}. In the GS structure, three pairs of carbon atoms and silicon atoms have the same distances due to the $C_{3v}$ symmetry. In the ES structure, all distances and bond lengths are increased because the $e$ defect orbitals are more delocalized that $a_1$ orbitals. The magnitude of the stretching is greater for carbon pairs than silicon pairs, which is consistent with the localization of defect orbitals. One pair of carbon atoms and silicon atoms has longer distance while the other two pairs have shorter distances, indicating that the symmetry is reduced from $C_{3v}$ to $C_{1h}$.

\begin{table}
  \caption{Distances (\AA) between atoms around the $kk$-VV$^0$ center in 4H-SiC in the equilibrium structures of the ground state (GS) and the excited state (ES). C1, C2 and C3 are three closet carbon atoms adjacent to the silicon-vacancy site while Si1, Si2 and Si3 are three closet silicon atoms adjacent to the carbon vacancy site, as shown in Fig.~\ref{s-fig:detailed-structure}.}
  \label{s-tab:detailed-kk-vv-dis}
  \begin{ruledtabular}
  \begin{tabular}{ccccccc}
    Atom pairs & \multicolumn{2}{c}{PBE} & \multicolumn{2}{c}{DDH} & \multicolumn{2}{c}{HSE}  \\
    & GS & ES & GS & ES & GS & ES \\
    \colrule
    C1$-$C2 &3.334 &3.456 &3.317 &3.448 &3.289 &3.424 \\
    C1$-$C3 &3.334 &3.465 &3.317 &3.457 &3.289 &3.433 \\
    C2$-$C3 &3.334 &3.456 &3.317 &3.448 &3.289 &3.424 \\
    Si1$-$Si2 &3.103 &3.114 &3.099 &3.108 &3.082 &3.088 \\
    Si1$-$Si3 &3.103 &3.147 &3.099 &3.140 &3.082 &3.120 \\
    Si2$-$Si3 &3.103 &3.114 &3.099 &3.108 &3.082 &3.088 \\
  \end{tabular}
  \end{ruledtabular}
\end{table}

\section{\label{s-sec:hybrid-phonon} Phonons at the level of hybrid functionals}
\begin{figure}
\includegraphics[width=8 cm]{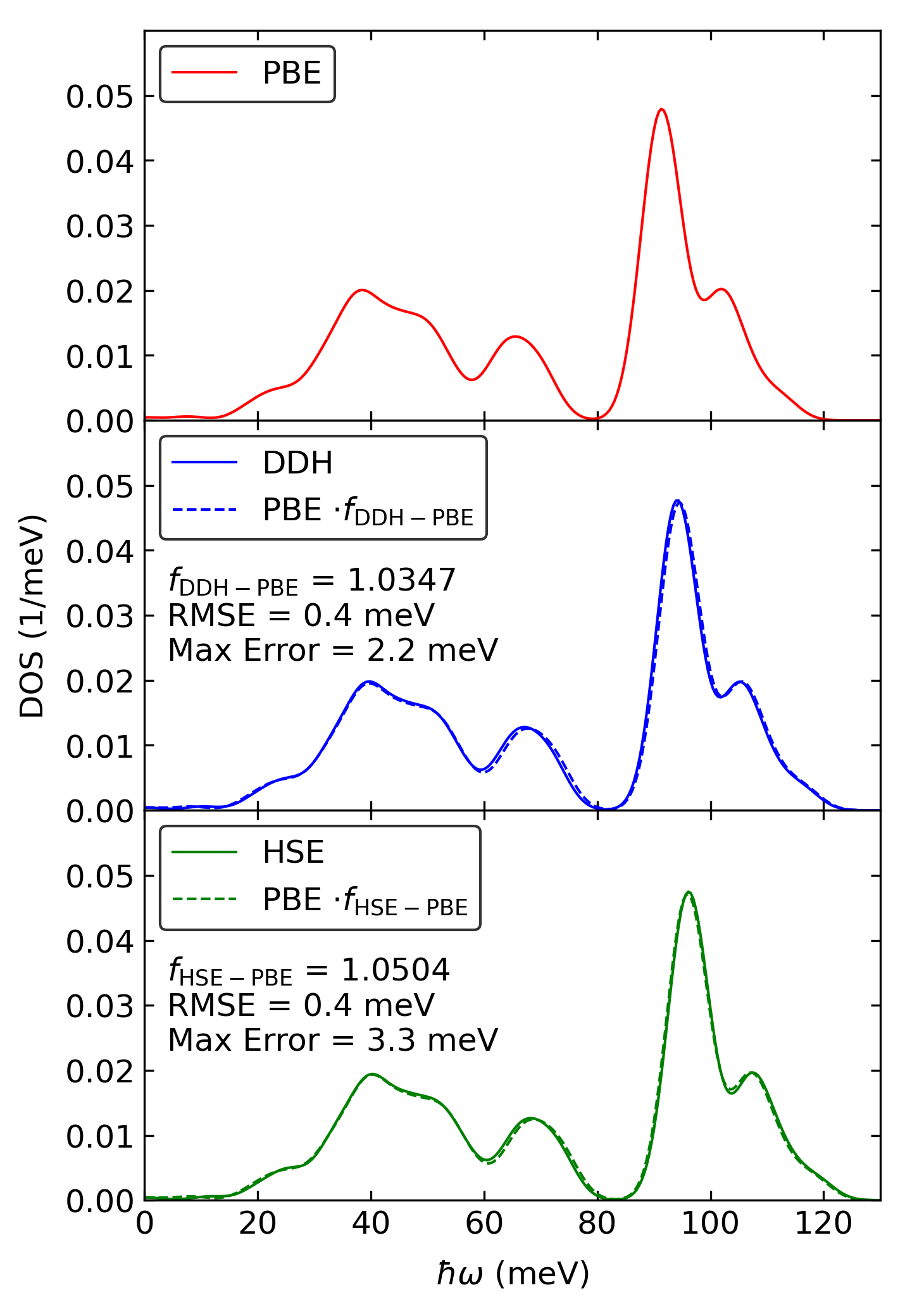}
\caption{\label{s-fig:vdos} Vibrational density of state of bulk 4H-SiC computed at the different levels of theory. Gaussian function with standard deviation $\sigma=3$ meV was used for broadening. Solid lines representing the results from direct first-principles calculations, and the dashed lines representing the results from approximation defined by Eq.~\ref{s-eqn: freq-ratio}.}
\end{figure}

For bulk diamond and 4H-SiC, phonons were calculated with PBE, DDH and HSE functionals using the $(4\times4\times4)$ supercell for diamond and and the $(5\times5\times2)$ supercell for 4H-SiC. Vibrational densities of states computed with different functionals are reported in Fig.~\ref{s-fig:vdos} for 4H-SiC.

Since it is computationally prohibitive to carry out phonon calculations for defect systems with hybrid functional, we first carried out phonon calculations at the PBE level of theory and then we assumed that the force constant matrix at the DDH or HSE level is just the one at PBE level multiplied by a constant. Under this hypothesis, the phonon eigenmodes obtained at the PBE level of theory would also diagonalize the dynamical matrix at the DDH/HSE level of theory, and the phonon frequencies at the DDH/HSE level of theory would be  proportional to the ones evaluated at PBE:
\begin{equation}
    \label{s-eqn: freq-ratio}
    \omega_{k,\mathrm{HSE}} = f_{\mathrm{HSE-PBE}} \cdot \omega_{k,\mathrm{PBE}},
\end{equation}
where $f$ is a scaling factor, evaluated as the ratio of the highest phonon frequency computed with DDH/HSE to that computed with PBE for the pristine systems (see Tab.~\ref{s-tab:freq-ratio}). The validity of the approximation was examined for bulk pristine systems by computing DDH/HSE phonon frequencies using Eq.~\ref{s-eqn: freq-ratio}. The maximum error, root mean square error (RMSE) and vibrational densities of state were displayed in Fig.~\ref{s-fig:vdos}. The overall agreement between vibrational density of state plots computed with Eq.~\ref{s-eqn: freq-ratio} is in very close agreement with those from direct first-principles calculations. The maximum error is about 2 to 3 meV, and the RMSE is less than 1 meV, validating the hypothesis. The same scaling factors were used to approximate the DDH/HSE level phonon frequencies for defect systems.

\begin{table}[b]
  \caption{Scaling factors of phonon frequencies at different level of theories defined in Eq.~\ref{s-eqn: freq-ratio} for diamond and 4H-SiC.}
  \label{s-tab:freq-ratio}
  \begin{ruledtabular}
    \begin{tabular}{ccc}
    & \multicolumn{1}{c}{$f_{\mathrm{DDH-PBE}}$} & \multicolumn{1}{c}{$f_{\mathrm{HSE-PBE}}$}  \\
    \hline
    Diamond &1.0400 &1.0514 \\
    4H-SiC &1.0347 &1.0504 \\
  \end{tabular}
  \end{ruledtabular}
\end{table}

\section{\label{s-sec:fc-embedding}Extrapolating to the dilute limit}

\begin{figure*}
\includegraphics[width=16 cm]{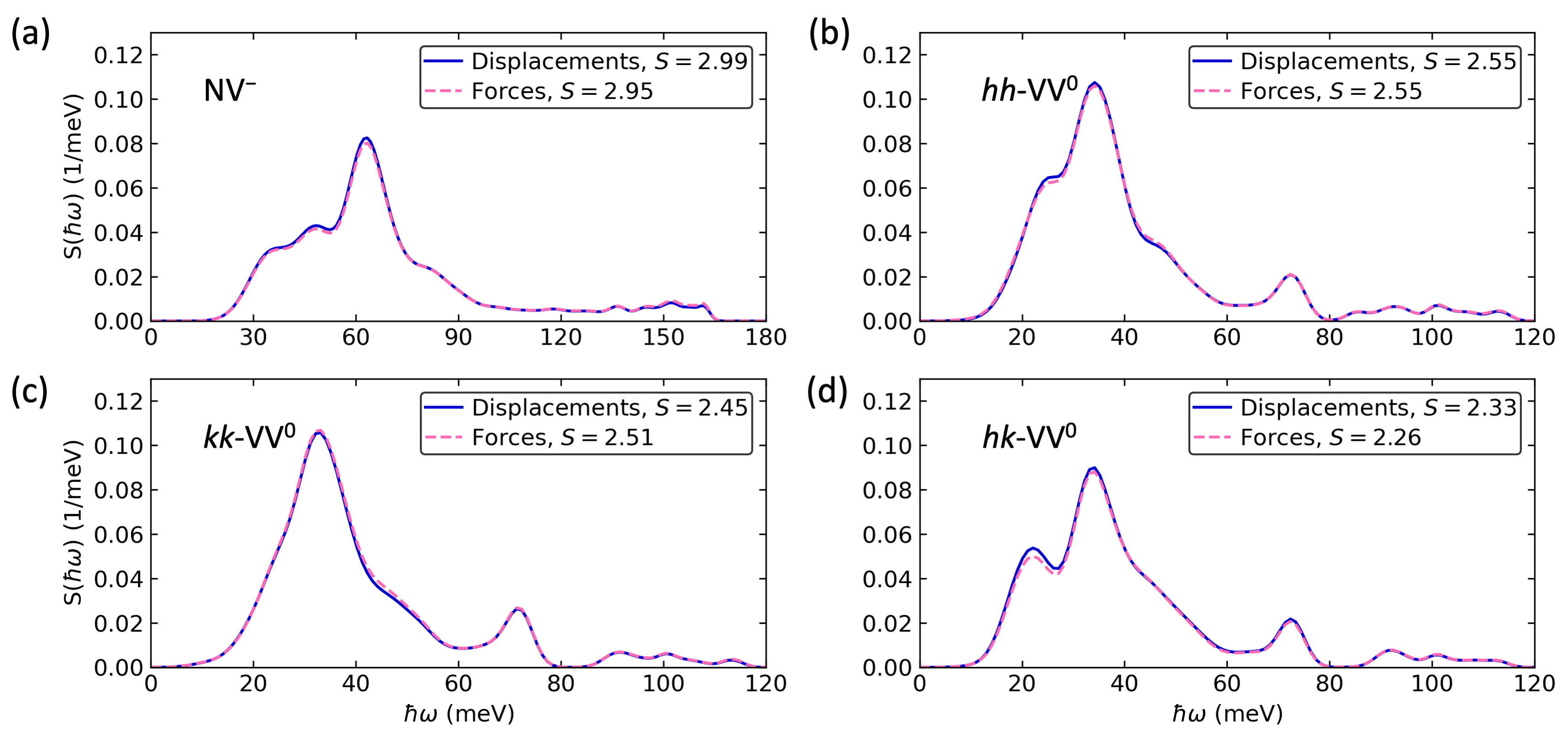}
\caption{\label{s-fig:dr-df-dq} Spectral densities of electron-phonon coupling for (a) NV$^-$ center in diamond, (b) $hh$-VV$^0$, (c) $kk$-VV$^0$, and (d) $hk$-VV$^0$ center in 4H-SiC. Calculations are carried out with the $(4\times4\times4)$ supercell for NV$^-$ and the $(5\times5\times2)$ supercell for VV$^0$ centers at the PBE level of theory. Solid blue lines denote results computed using Eq.~\ref{s-eqn:dq-displacement}, and dashed red lines denote results computed using Eq.~\ref{s-eqn:dq-force}. Total Huang-Rhys factors (HRFs, $S$) are also given in the legend.}
\end{figure*}

Mass-weighted atomic displacements of the $k$-th phonon mode, $\Delta Q_k$, are required in order to compute HRFs. Either the atomic displacements computed between the equilibrium structures of the GS and ES (Eq.~\ref{s-eqn:dq-displacement}) or GS forces evaluated at the equilibrium structure of the ES (Eq.~\ref{s-eqn:dq-force}) can be used to compute $\Delta Q_k$~\cite{Alkauskas_2014, razinkovas2020vibrational}.
\begin{equation}
\label{s-eqn:dq-displacement}
    \Delta Q_{k}=\sum_{\alpha=1}^N\sum_{i=x,y,z}  \sqrt{M_{\alpha}} \Delta \mathbf{R}_{\alpha i} \boldsymbol{e}_{k,\alpha i},
\end{equation}
\begin{equation}
\label{s-eqn:dq-force}
    \Delta Q_{k}=\frac{1}{\omega_{k}^{2}} \sum_{\alpha=1}^N\sum_{i=x,y,z} \frac{\mathbf{F}_{\alpha i}}{\sqrt{M_{\alpha}}} \boldsymbol{e}_{k,\alpha i}.
\end{equation}
Here $\mathbf{e}_{k,\alpha i}$ is the eigenvector of the $k$-th phonon mode on the $\alpha$-th atom in the $i$-th direction. $M_\alpha$ is the mass of the $\alpha$-th atom, and $\Delta \mathbf{R}_{\alpha i} = (\mathbf{R}_{\alpha i})_e - (\mathbf{R}_{\alpha i})_g$ is the displacement between the ES and the GS equilibrium atomic structures in the $i$-th direction. $\mathbf{F}_{\alpha i}$ is the GS force on the $\alpha$-th atom in the $i$-th direction evaluated at the ES equilibrium structure. These two approaches are equivalent under the harmonic approximation:
\begin{equation}
    \Delta \mathbf{F} = \mathbf{H} \cdot \Delta \mathbf{R}.
\end{equation}
Here $\mathbf{H}$ is the Hessian matrix or the force constant matrix. We used both approaches to compute HRFs and spectral densities of electron-phonon coupling for the NV$^-$ center in diamond with the $(4\times4\times4)$ supercell and the VV$^0$ centers with $(5\times5\times2)$ supercells, as shown in Fig.~\ref{s-fig:dr-df-dq}. The differences in HRFs computed with the two approaches are within 3\%, and the spectral densities are almost identical, indicating that the harmonic approximation works well for these defect systems and validating the use of Eq.~\ref{s-eqn:dq-force} in our work.

\begin{figure}
\includegraphics[width=8cm]{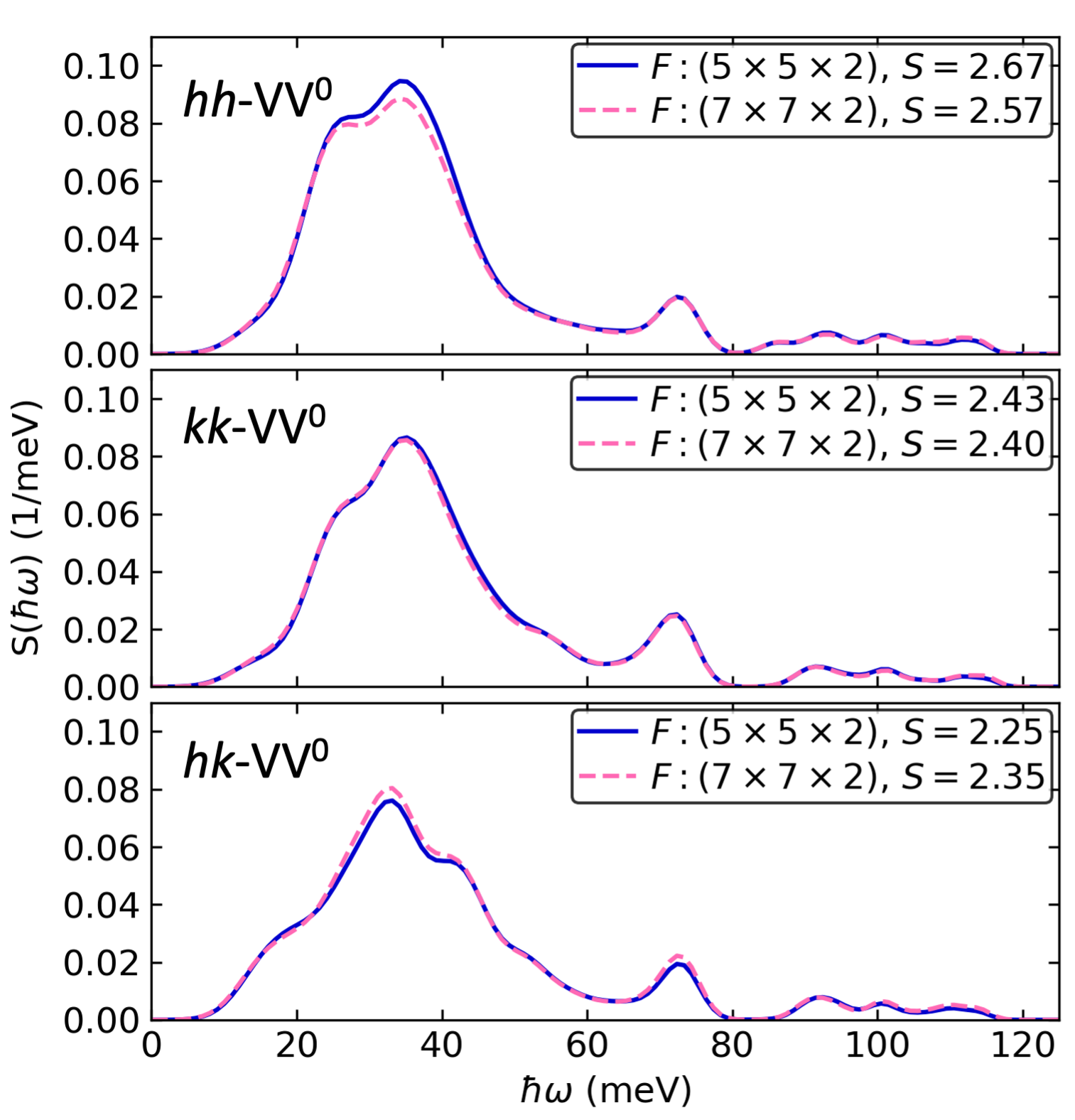}
\caption{\label{s-fig:k-vv-400-784} Spectral densities of electron-phonon coupling for $hh$-VV$^0$, $kk$-VV$^0$ and $hk$-VV$^0$ centers in 4H-SiC for a $(7 \times 7 \times 2)$ supercell. Calculations are carried out using either the forces computed with a $(5 \times 5 \times 2)$ supercell (solid blue lines) or a $(7 \times 7 \times 2)$ supercell (dashed red lines) at the PBE level of theory. Total Huang-Rhys factors (HRFs, $S$) are also given in the legend.}
\end{figure}

In order to compute HRFs and spectral densities for supercells larger than $(4\times4\times4)$ for NV$^-$ and $(5\times5\times2)$ VV$^0$ centers, both (i) $\Delta Q_k$ and (ii) phonons for these supercells are needed. For (i) $\Delta Q_k$, since direct first principles calculations with these supercells are prohibitive, we used the hypothesis that the forces quickly decay to zero as a function of the distance to the defect center, and the forces computed with the smallest supercell ($(4\times4\times4)$ for NV$^-$ and $(5\times5\times2)$ VV$^0$ centers) are the same as those computed with the larger supercell. A previous work shows that forces 5 \AA{} away from the defect center contribute negligibly to the HRF for the NV$^-$ in diamond~\cite{razinkovas2020vibrational}. Here, we want to ensure that for VV$^0$ centers in 4H-SiC, forces computed with the smallest $(5 \times 5 \times 2)$ supercell are sufficient to represent the forces in larger supercells. For this purpose, $(7 \times 7 \times 2)$ supercells were used to model the VV$^0$ centers. 
HRFs and spectral densities were computed using forces computed with $(7 \times 7 \times 2)$ supercells and compared with those computed using forces computed with $(5 \times 5 \times 2)$ supercells, as shown in Fig.~\ref{s-fig:k-vv-400-784}. The differences in HRFs are less than 5\%, and the spectral densities are almost identical, indicating that forces computed using $(5 \times 5 \times 2)$ supercells are sufficient to evaluate $\Delta Q_k$ for larger supercells.

We used the force constant matrix embedding scheme proposed by Alkauskas et al.~\cite{Alkauskas_2014, SiV_Phonon_2018, razinkovas2020vibrational} to compute (ii) phonons for supercells larger than $(4\times4\times4)$ for the NV$^-$ center in diamond and $(5\times5\times2)$ for VV$^0$ centers in 4H-SiC. The method is based on the short-range property of the force constant matrix in semiconductors: when the position of one atom changes in a fixed electronic state, the induced force on neighboring atoms decays rapidly to zero as a function of the distance from this atom. It enables the construction of the force constant matrix of larger supercells using the one computed with the smallest supercells. The force constant matrix is defined as
\begin{equation}
    \Phi_{\alpha, \beta}(i, j) = \frac{\partial F_{i, \alpha}}{\partial r_{j, \beta}},
\end{equation}
where $F_{i,\alpha}$ is the force that acts on atom $i$ in the Cartesian direction $\alpha$ and $r_{j,\beta}$ is the displacement of atom $j$ from the equilibrium position in the direction $\beta$. The force constant matrix of a large defect supercell is constructed as follows. If atoms $n$ and $m$ are separated by a distance larger than a chosen cutoff radius $r_{c1}$, then the force constant matrix element is set to zero. If both atoms are separated from the defect center by a distance smaller than the cutoff radius $r_{c2}$, then the force constant matrix element from the actual defect supercell is used. For all other atom pairs the force constant matrix elements of the bulk system are used. To fulfill the acoustic sum rule, we use the same approach as the one used in Ref~\cite{razinkovas2020vibrational}:
\begin{equation}
    \Phi_{\alpha, \alpha}(n, n)=-\sum_{m \neq n} \Phi_{\alpha, \alpha}(m, n).
\end{equation}

\begin{figure*}
    \includegraphics[width=16 cm]{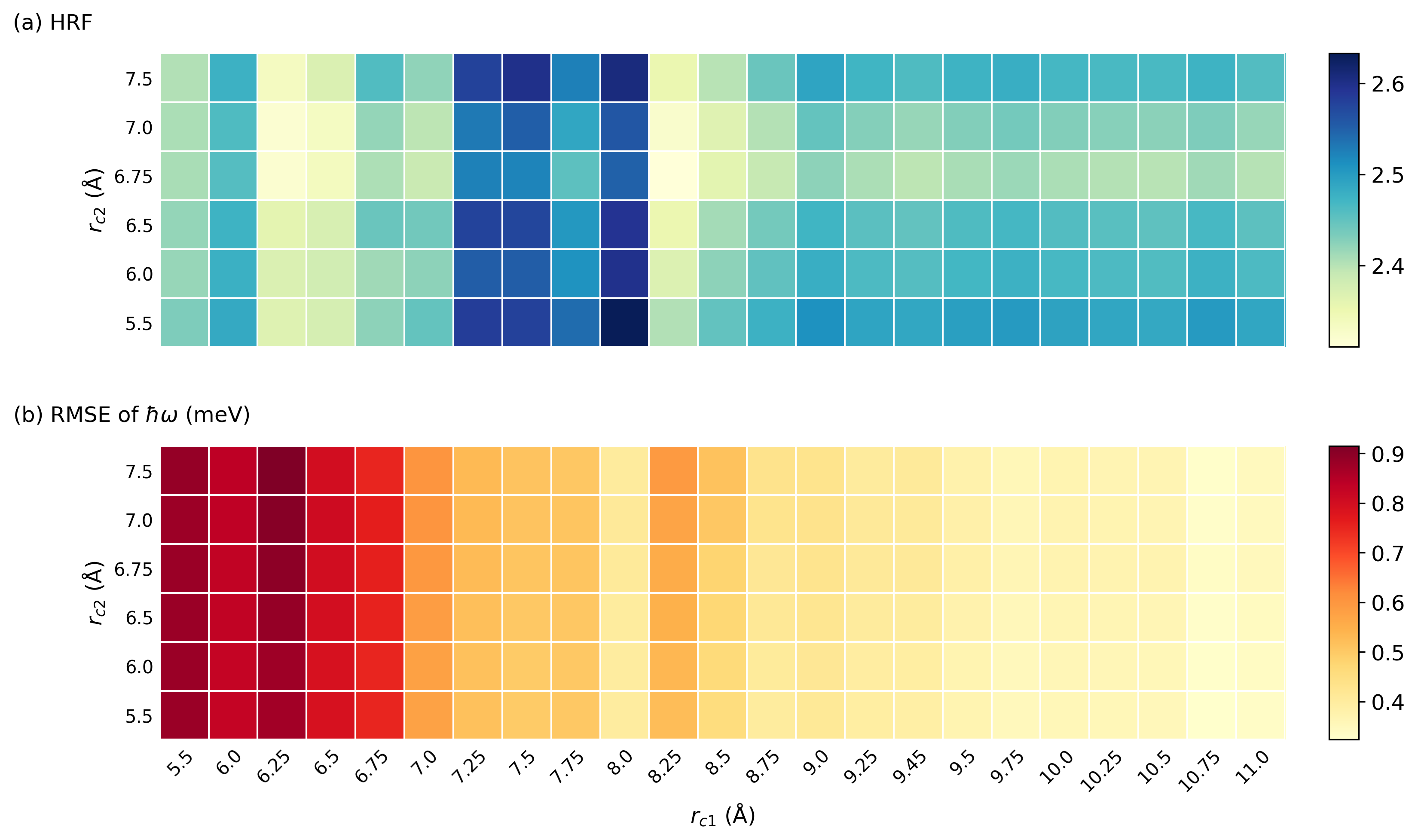}
    \caption{\label{s-fig:kk-vv-cutoff-radius} (a) Huang-Rhys factor (HRF) as a function of the cutoff radius $r_{c1}$ and $r_{c2}$. (b) Root mean square error (RMSE) of phonon energies $\hbar\omega$ with those computed from the first-principles calculations as a function of the cutoff radius $r_{c1}$ and $r_{c2}$. $(5\times 5\times 2)$ supercell was used for these calculations. Forces and force constant matrix computed at the PBE level of theory were used.}
\end{figure*}

To obtain components for the embedding process, $(4 \times 4 \times 4)$ supercells were used to compute the force constant matrix for both pristine bulk diamond and the NV$^-$ center in diamond, and $(5 \times 5 \times 2)$ supercells were used to compute the force constant matrix for VV$^0$ centers in 4H-SiC, and a $(8 \times 8 \times 3)$ supercell with 1536 sites were used to compute the force constant matrix for pristine bulk 4H-SiC. As for the cutoff radius, $r_{c1} = 5$ \AA{} and $r_{c2}=5$ \AA{} was used for the NV$^-$ in diamond and $r_{c1} = 9.45$ \AA{} and $r_{c2}=6.75$ \AA{} was used for VV$^0$ centers in 4H-SiC. Choice of cutoff radii was carefully examined. Taking the $kk$-VV$^0$ center in 4H-SiC modeled by the $(5\times 5\times 2)$ supercell as the example, HRF and the error of the phonon energies are computed using the force constant matrix from the embedding scheme with different choice of cutoff radius, as shown in Fig.~\ref{s-fig:kk-vv-cutoff-radius}. It can be concluded that $r_{c1} = 9.45$ \AA{} and $r_{c2}=6.75$ \AA{} is a good choice: the HRF is only 5\% away from the reference value, and the root mean square error (RMSE) of phonon energies is 0.4 meV. Test on larger supercells points out the existence of an uncertainty of $\sim$0.2 in the HRF due to the choice of $r_{c1}$, and we only kept 1 digit for these HRFs.

\begin{figure}
\includegraphics[width=8 cm]{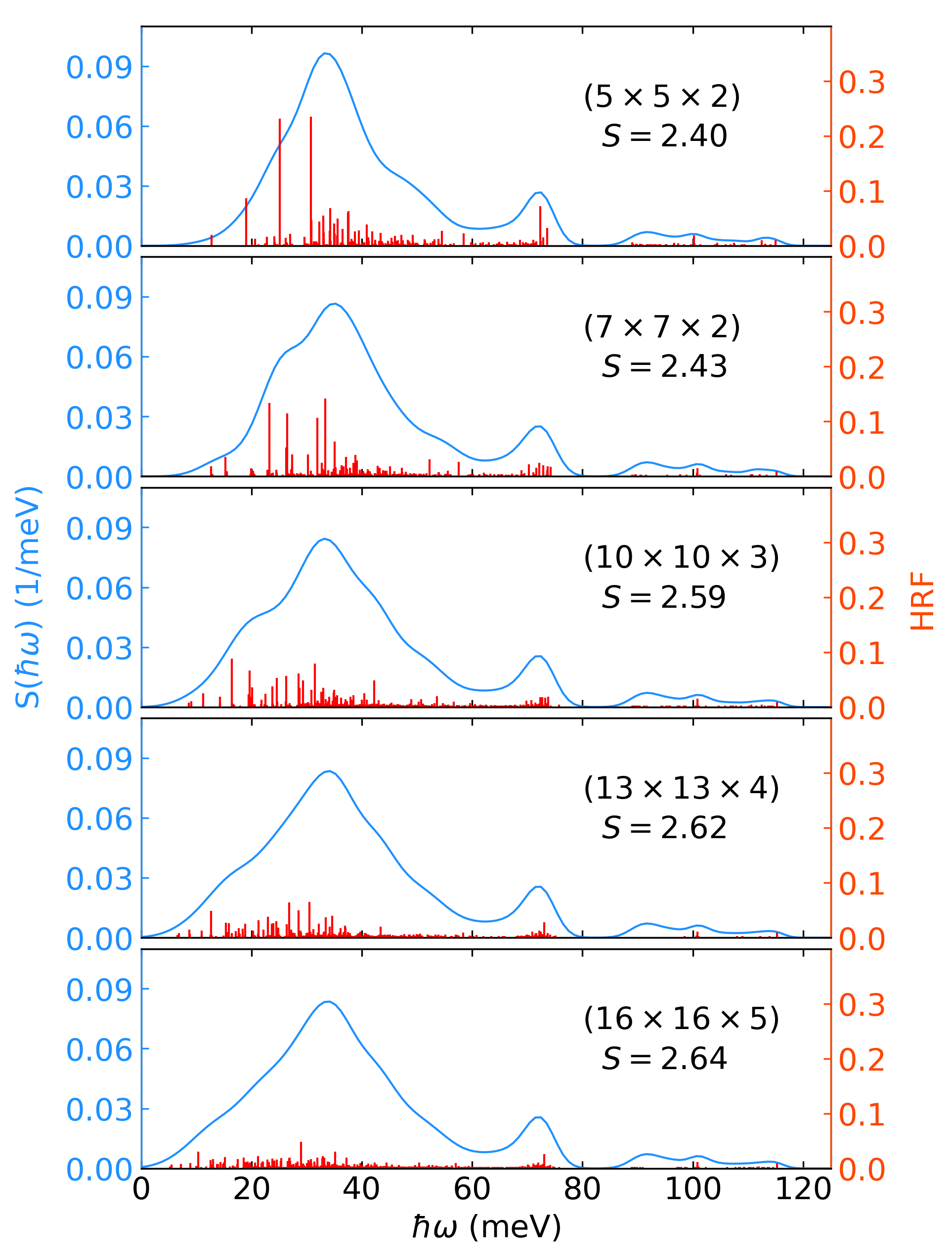}
\caption{\label{s-fig:kk-VV-convergence} Convergence of the spectral density of electron-phonon coupling $S(\hbar\omega)$ and Huang-Rhys factors (HRFs) with respect to the supercell size for the $kk$-VV$^0$ center in 4H-SiC computed at the PBE level of theory. Supercells range in size from $(5 \times 5 \times 2)$ (400 atomic sites) to $(16 \times 16 \times 5)$ (10240 atomic sites). Blue lines refer to spectral densities, and the red vertical bars represent the partial HRFs for each phonon modes.}
\end{figure}

We examined the convergence of partial HRFs and the spectral density as a function of the supercell size for the $kk$-VV$^0$ center in 4H-SiC, as shown in Fig.~\ref{s-fig:kk-VV-convergence}. By extrapolating to the dilute limit, modes at 23, 33, and 72 meV split into many closely spaced modes, with a simultaneous decrease of their absolute contributions to the total HRF, indicating the existence of quasi-local vibrational modes.

We also examined the computed highest and lowest phonon energy for different supercells. At the PBE level of theory we find that the lowest phonon energy is 34 meV (11 meV) for the NV$^-$ center in diamond when the $(4\times4\times4)$ ($(12\times12\times12)$) supercell is used. The lowest phonon energy is 11 meV (5 meV) for VV$^0$ centers in 4H-SiC when the $(5\times5\times2)$ ($(16\times16\times5)$) supercell is used. The computed highest phonon energy depends weakly on the supercell size. At the PBE level theory we find that the highest phonon energy is 162$\sim$163 meV for the NV$^-$ center in diamond, basically the same as the highest bulk phonon energy, 163 meV. The highest phonon energy is 115 meV for VV$^0$ centers in 4H-SiC, very close to that of the highest bulk phonon energy, 114 meV.

\section{\label{s-sec:vibrational-modes-ana} Detailed analysis of spectral densities and vibrational modes}

The spectral density of the NV$^-$ center in diamond is dominated by a peak at about 63 meV, which originates from the coupling with a quasi-local vibrational mode~\cite{Alkauskas_2014}. Coupling with other (quasi-)local vibrational modes leads to detailed structures above 130 meV. In the case of VV$^0$ centers, our calculations showed two peaks at about 34 meV and 72 meV together with detailed structures above 90 meV. The overall intensity of $hh$-VV$^0$ is larger than that of $kk$-VV$^0$, which is in turn larger than that of $hk$-VV$^0$, consistent with the magnitude of the HRF. We examined the convergence of the HRF and the spectral densities as a function of the supercell size for $kk$-VV$^0$ (see Fig.~\ref{s-fig:kk-VV-convergence}). By extrapolating to the dilute limit, we found that modes at 23, 33, and 72 meV split into many closely spaced ones, with a simultaneous decrease of their absolute contributions, indicating the existence of quasi-local vibrational modes. The same behavior was also observed for $hh$-VV$^0$ and $hk$-VV$^0$. Spectral densities of $hh$-VV$^0$ and $kk$-VV$^0$ are similar, with two small differences: (i) the shoulder peak at 23 meV is more pronounced for $hh$-VV$^0$; (ii) $hh$-VV$^0$ has a small peak at about 86 meV not present for $kk$-VV$^0$. The $hk$-VV$^0$ spectrum shows an apparent shoulder peak at 22 meV.

\begin{figure*}
\includegraphics[width=16 cm]{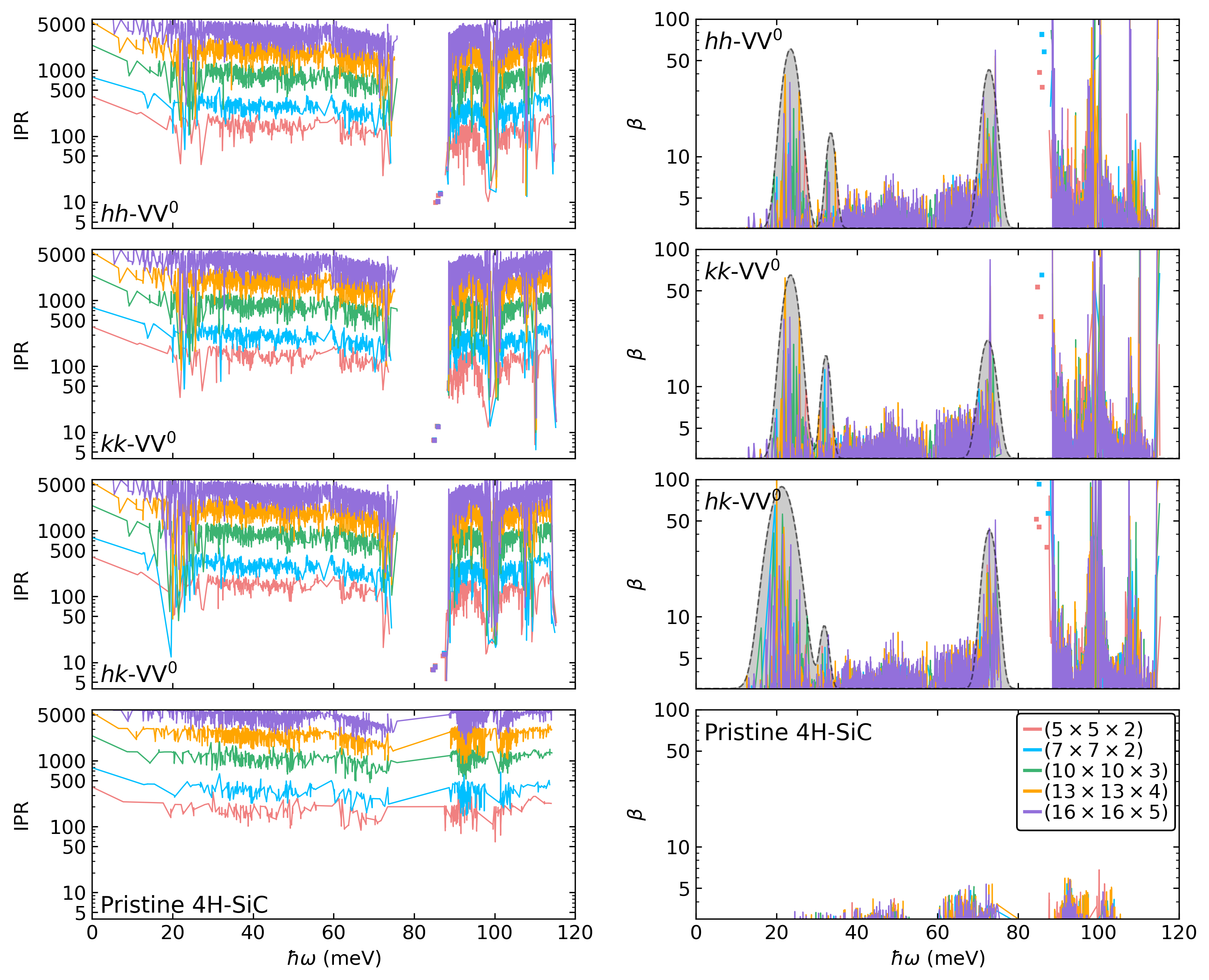}
\caption{\label{s-fig:IPR-beta-hh-hk-vv} Inverse partition ratio (IPR) and localization ratio ($\beta$) as a function of phonon energies computed using supercells of different sizes for $hh$-VV$^0$, $kk$-VV$^0$, and $hk$-VV$^0$ centers in 4H-SiC and pristine 4H-SiC. The shaded region denotes the quasi-local vibrational modes.}
\end{figure*}

\begin{figure*}
\includegraphics[width=16 cm]{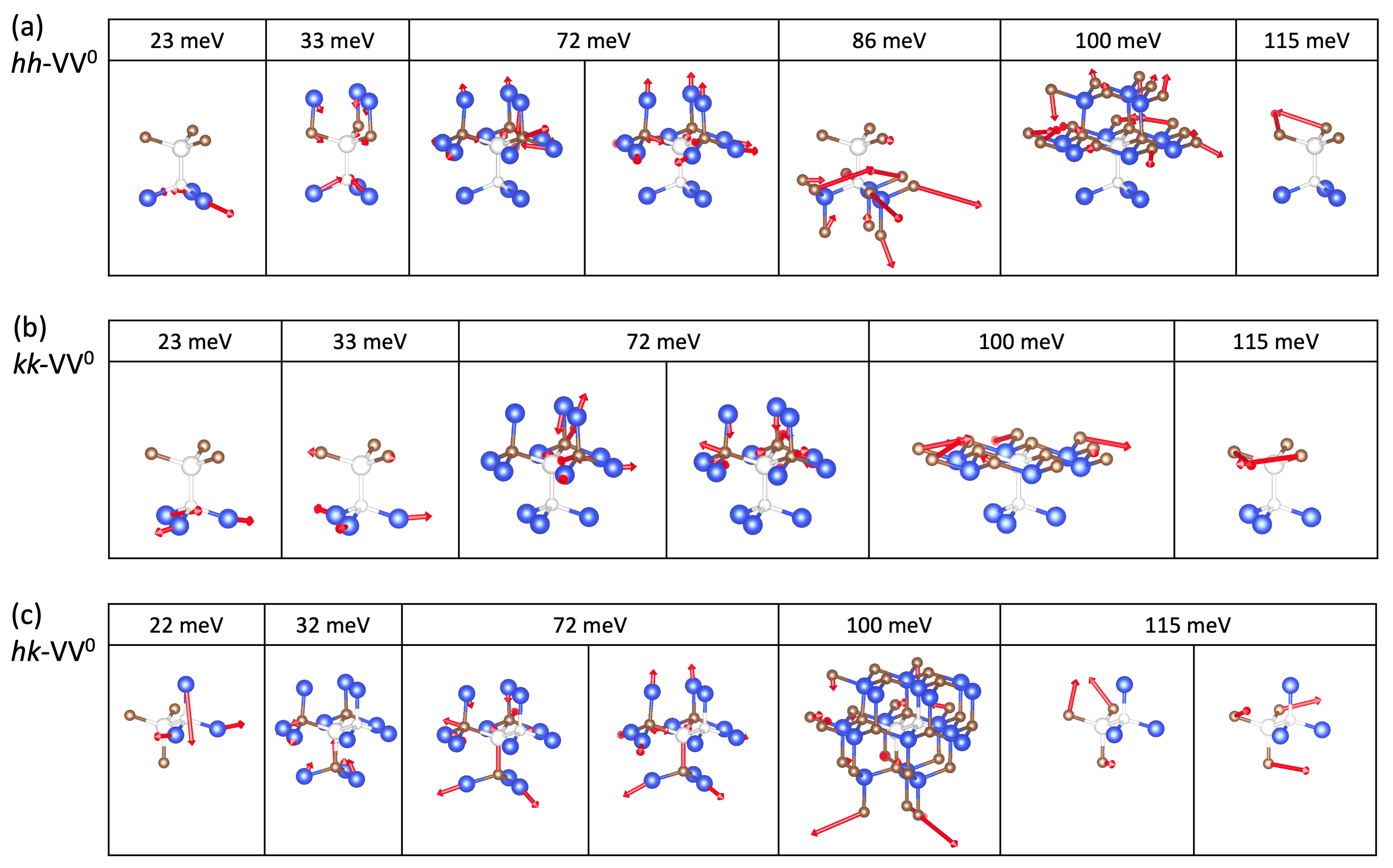}
\caption{\label{s-fig:hh-hk-vibrations} Displacement patterns of vibrational modes of (a) $hh$-VV$^0$, (b) $kk$-VV$^0$ and (c) $hk$-VV$^0$ centers in 4H-SiC. Vectors are amplified by a factor of 10. Vibrational modes with energy smaller than 80 meV are quasi-local modes while those with energy greater than 80 meV are local modes.}
\end{figure*}

We performed phonon (vibrational) modes analysis for VV$^0$ centers in 4H-SiC to comprehend their relationship with the defect center using the same approach as the one used to study the NV$^-$ and the SiV$^-$ centers in diamond by Alkauskas et al.~\cite{Alkauskas_2014, SiV_Phonon_2018}. The quasi-local and local vibrational modes were qualitatively characterized by computing the inverse partition ratio (IPR) and localization ratio $\beta_k$. the IPR for the $k$-th vibrational modes is defined as
\begin{equation}
    \mathrm{IPR}_k =\dfrac{1}{\sum_{\alpha=1}^N \left(\sum_{i=x,y,z} \mathbf{e}^2_{k,\alpha i}\right)^2}.
\end{equation}
The IPR reflects the effective number of atoms that participate in a phonon mode; $\text{IPR} = 1$ indicates that only one atom vibrates, while $\text{IPR} = N$ indicates all $N$ atoms in the supercell vibrate with the same amplitude. $\beta_k$ is defined as
\begin{equation}
    \beta_k = \dfrac{N}{\mathrm{IPR}_k},
\end{equation}
and describes the inverse fraction of atoms in the supercell that vibrate for a given phonon mode; $\beta_k \gg 1$ for quasi-local and local modes. Here we analyzed the vibrational modes for $hh$-VV$^0$, $kk$-VV$^0$ and $hk$-VV$^0$ centers in 4H-SiC by computing IPR and $\beta_k$ for different supercell sizes, as shown in Fig.~\ref{s-fig:IPR-beta-hh-hk-vv} together with results for the pristine 4H-SiC. Three inverted peaks with the energy $\sim$23 meV, 33 meV, and 72 meV can be identified in the IPR plots with IPRs significantly smaller than those for the pristine 4H-SiC. IPRs for these peaks increase as the supercell size increases, which is a signature of quasi-local modes. The quasi-local modes can also be identified in the $\beta$ plots as isolated peaks. These quasi-local modes are made of a continuum of vibrations and have significant localization on the atoms around the defect center (see Fig.~\ref{s-fig:hh-hk-vibrations}). The 23 meV quasi-local mode involves asymmetric stretching of three nearest neighbor silicon atoms around $\mathrm{V}_{\mathrm{C}}$.
The 33 meV quasi-local mode involves symmetric stretching of three nearest neighbor silicon atoms around $\mathrm{V}_{\mathrm{C}}$.
The 72 meV mode involves both symmetric and asymmetric vibrations of three nearest neighbor carbon atoms and nine next nearest neighbor silicon atoms around $\mathrm{V}_{\mathrm{Si}}$.
Several other inverted peaks with energy at 86 meV, 100 meV, 110 meV, and 115 meV can be identified in the IPR plots with IPRs around 10, indicating the existence of local vibrational modes. Their contributions to the PL line shape are much smaller compared with the quasi-local vibrational modes.

With the knowledge of vibrational modes, we can interpret the observed difference among the spectral densities of VV$^0$ centers in 4H-SiC. 
The 23 meV peak in the spectral density originates from the coupling with the mode involving asymmetric vibrations of three nearest neighbor silicon atoms around $\mathrm{V}_{\mathrm{C}}$ where two silicon atoms vibrate with an amplitude smaller than the other silicon atom. We computed the difference of mass-weighted displacements among these silicon atoms and obtained 0.15 amu$^{0.5}$ \AA{} for $hh$-VV$^0$ and 0.07 amu$^{0.5}$ \AA{} for $kk$-VV$^0$. This fact indicates that the coupling with the 23 meV mode is stronger for $hh$-VV$^0$ and explains the observation that the 23 meV shoulder peak is more notable in the spectral densities for $hh$-VV$^0$ than $kk$-VV$^0$. For $hk$-VV$^0$, the 22 meV mode is more localized with more considerable variance compared with $hh$-VV$^0$ and $kk$-VV$^0$, as can be seen from the figure of $\beta$. It is consistent with the more prominent feature of the 22 meV shoulder peak in the spectral density.

\section{Calculations of transition dipole moment and radiative lifetime}

In order to validate the Franck-Condon (FC) approximation using the 1D model, we computed the electronic transition dipole moment between the ES and the GS, $|\boldsymbol{\mu}_{eg}|$, for the actual defect systems along the 1D configuration coordinate. By approximating the optical transition from the ES to the GS as a transition between  single-particle states, we have that
\begin{equation}
    \boldsymbol{\mu}_{eg} = - \frac{e\hbar^2}{(\varepsilon_f - \varepsilon_i)m} \langle \psi_f | \boldsymbol{\nabla} | \psi_i \rangle,
\end{equation}
where $e$ is the charge of the electron, $m$ is the mass of the electron, $\hbar$ is the Planck constant, $\psi_f$ and $\psi_i$ are Kohn-Sham orbitals of defect levels involved in the optical transition, and $\varepsilon_f$ and $\varepsilon_i$ are the corresponding energies. To validate the correctness of the calculation, we compared the radiative lifetime $\tau_{\text{rad}}$ at the PBE level of theory. $\tau_{\text{rad}}$ is related to the radiative emission rate $\Gamma_{\text{rad}}$ by
\begin{equation}
    \Gamma_{\text{rad}} = \dfrac{1}{\tau_{\text{rad}}} = \dfrac{nE_{\text{ZPL}}^3|\boldsymbol{\mu}_{eg}|^2}{3\pi\epsilon_0c^3\hbar^4},
\end{equation}
where $n=2.4$ and $n=2.6473$ are the refractive index of diamond and 4H-SiC, respectively. $E_{\text{ZPL}}=1.945$ eV and $E_{\text{ZPL}}=1.096$ eV are the experimental ZPL energies of the NV$^-$ center in diamond and the $kk$-VV$^0$ center in 4H-SiC, respectively. $\epsilon_0$ is the vacuum permittivity, and $c$ is the speed of light in vacuum. For the NV$^-$ in diamond, the computed lifetime is 12.4 ns (10.3 ns) when the equilibrium atomic structure of GS (ES) is used. It is in good agreement with a previous theoretical result 12.2 ns~\cite{razinkovas2021photoionization} and the experimental result 12 ns~\cite{doherty2013nitrogen}. For the $kk$-VV$^0$ in 4H-SiC, the computed lifetime is 36.8 ns (31.4 ns) when the equilibrium atomic structure of the GS (ES) is used. It is in good agreement with a previous theoretical result 38.49 ns~\cite{Davidsson2020lifetime}.

\begin{figure}
    \centering
    \includegraphics[width=9cm]{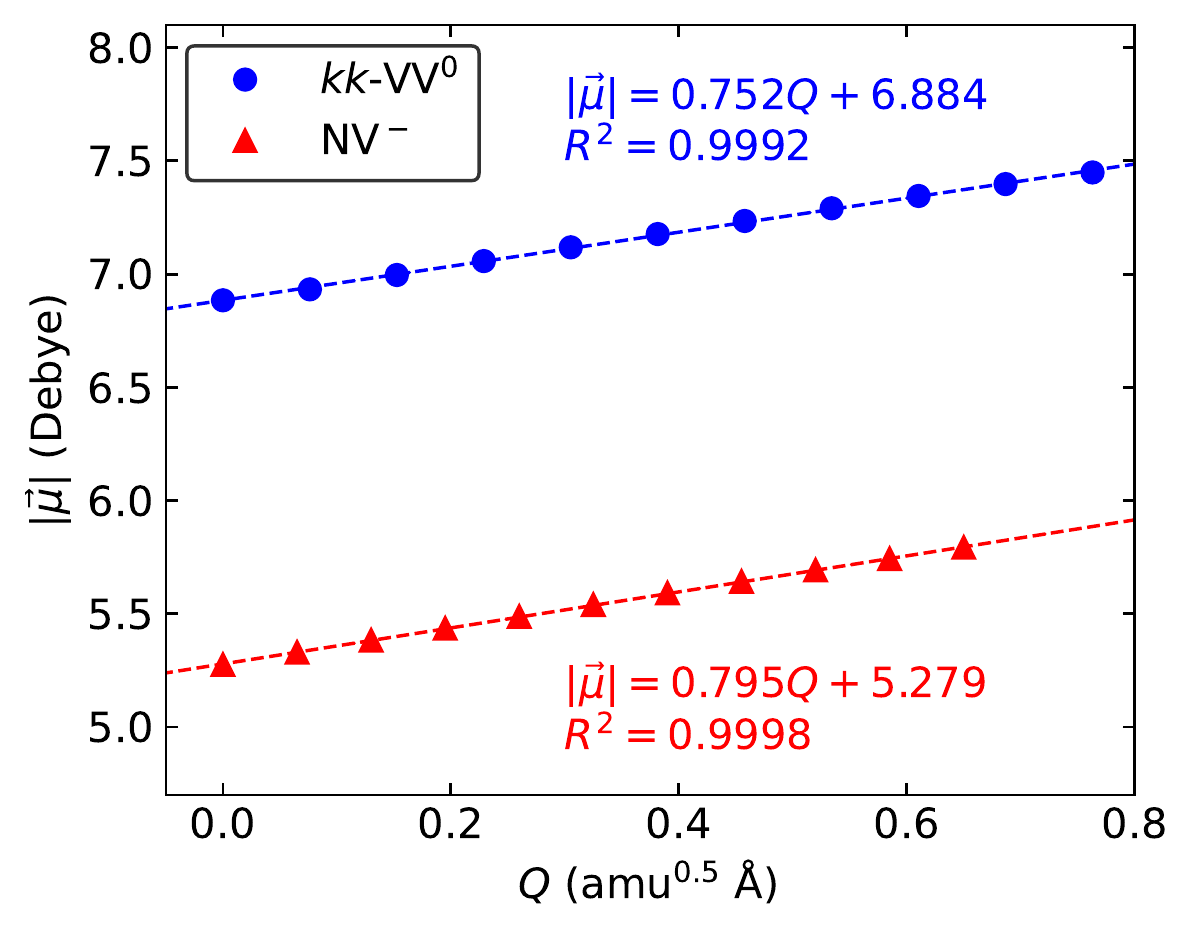}
    \caption{Norm of the calculated transition dipole moment $|\boldsymbol{\mu}|$ between the ground state and the excited state as a function of the configuration coordinate $Q$ for the NV$^-$ center in diamond and the $kk$-VV$^0$ center in 4H-SiC. Calculations are performed at the PBE level of theory with a $(4\times 4\times 4)$ supercell for the NV$^-$ center in diamond and a $(7\times 7\times 2)$ supercell for the $kk$-VV$^0$ center in 4H-SiC. The parameters of the linear fits are reported in the figure.}
    \label{s-fig:TDM}
\end{figure}

We have numerically computed the electronic transition dipole moment $|\boldsymbol{\mu}|$ as a function of the configuration coordinate $Q$ for the NV$^-$ center in diamond and the $kk$-VV$^0$ center in 4H-SiC, as shown in Fig.~\ref{s-fig:TDM}. A linear dependence with a relative change of about 10\% between the equilibrium atomic structures of GS and ES can be found. The derivative of $|\boldsymbol{\mu}|$ with respect to $Q$ is then used to compute the Franck-Condon Herzberg-Teller (FCHT) and Herzberg-Teller (HT) terms of the PL line shape using the one-dimensional (1D) model:
\begin{equation}
    L_{\mathrm{FC}}\left(\hbar\omega,T\right)\propto\omega^3\sum_{i}\sum_{j}{P_{ej}\left(T\right)}\left|\boldsymbol{\mu}_{eg}^0\right|^2\left|\left\langle\phi_{n^{ej}}\middle|\phi_{n^{gi}}\right\rangle\right|^2\delta\left(E_{\mathrm{ZPL}}+E_{ej}-E_{gi}-\hbar\omega\right),
\end{equation}
\begin{equation}
\begin{aligned}
    L_{\mathrm{FCHT}}\left(\hbar\omega,T\right)&\propto{2\omega}^3\sum_{i}\sum_{j}{P_{ej}\left(T\right)}\boldsymbol{\mu}_{eg}^0\frac{d\boldsymbol{\mu}_{eg}^0}{dQ}\left\langle\phi_{n^{ej}}\middle|\phi_{n^{gi}}\right\rangle\left\langle\phi_{n^{gi}}\middle| Q\middle|\phi_{n^{ej}}\right\rangle\\
    &\quad \times \delta\left(E_{\mathrm{ZPL}}+E_{ej}-E_{gi}-\hbar\omega\right),
\end{aligned}
\end{equation}
\begin{equation}
    L_{\mathrm{HT}}\left(\hbar\omega,T\right)\propto\omega^3\sum_{i}\sum_{j}{P_{ej}\left(T\right)}\left|\frac{d\boldsymbol{\mu}_{eg}^0}{dQ}\right|^2\left|\left\langle\phi_{n^{gi}}\middle| Q\middle|\phi_{n^{ej}}\right\rangle\right|^2\delta\left(E_{\mathrm{ZPL}}+E_{ej}-E_{gi}-\hbar\omega\right).
\end{equation}
Here $\boldsymbol{\mu}_{eg}^0$ is $\boldsymbol{\mu}_{eg}$ at $Q=0$, and $\frac{d\boldsymbol{\mu}_{eg}^0}{dQ}$ is the derivative of $\boldsymbol{\mu}_{eg}$ at $Q=0$.

\bibliographystyle{apsrev4-2}
\bibliography{si}